\documentclass[iop,apj]{emulateapj}
\usepackage{apjfonts}
\usepackage{graphicx}

\shorttitle{Isotropy constraints on powerful sources of ultrahigh-energy cosmic rays at $10^{19}$ eV}
\shortauthors{Takami, Murase, \& Dermer}

\begin{document}


\title{Isotropy constraints on powerful sources of ultrahigh-energy cosmic rays at $10^{19}$ eV}

\author{Hajime Takami\altaffilmark{1,5}}
\author{Kohta Murase\altaffilmark{2,3,6}}
\author{Charles D. Dermer\altaffilmark{4}}

\altaffiltext{1}{Theory Center, Institute of Particle and Nuclear Studies, High Energy Accelerator Research Organization (KEK), 1-1, Oho, Tsukuba, Ibaraki 305-0801, Japan}
\altaffiltext{2}{Institute for Advanced Study, 1 Einstein Dr. Princeton, New Jersey 08540, USA}
\altaffiltext{3}{Center for Particle and Gravitational Astrophysics; Department of Physics; Department of Astronomy \& Astrophysics, The Pennsylvania State University, University Park, Pennsylvania 16802, USA}
\altaffiltext{4}{Code 7653, Space Science Division, U.S. Naval Research Laboratory, Washington, DC 20375, USA}
\altaffiltext{5}{JSPS Research Fellow}
\altaffiltext{6}{Hubble Fellow}

\begin{abstract}
Anisotropy in the arrival direction distribution of ultrahigh-energy cosmic rays (UHECRs) produced by powerful sources is numerically evaluated. We show that, taking account of the Galactic magnetic field, nondetection of significant anisotropy at $\approx 10^{19}$ eV at present and in future experiments imposes general upper limits on UHECR proton luminosity of steady sources as a function of source redshifts. The upper limits constrain the existence of typical steady sources in the local universe and limit the local density of $10^{19}$ eV UHECR sources to be $\gtrsim 10^{-3}$ Mpc$^{-3},$ assuming average intergalactic magnetic fields less than $10^{-9}$ G. This isotropy, which is stronger than measured at the highest energies, may indicate the transient generation of UHECRs. Our anisotropy calculations are applied for extreme high-frequency-peaked BL Lac objects 1ES 0229+200, 1ES 1101-232, and 1ES 0347-121, to test the UHECR-induced cascade model, in which beamed UHECR protons generate TeV radiation in transit from sources. While the magnetic-field structure surrounding the sources affects the required absolute cosmic-ray luminosity of the blazars, the magnetic-field structure surrounding the Milky Way directly affects the observed anisotropy.  If both of the magnetic fields are weak enough, significant UHECR anisotropy from these blazars should be detectable by the Pierre Auger Observatory unless the maximum energy of UHECR protons is well below $10^{19}$~eV. Furthermore, if these are the sources of UHECRs above $10^{19}$ eV, a local magnetic structure surrounding the Milky Way is needed to explain the observed isotropy at $\sim 10^{19}$ eV, which may be incompatible with large magnetic structures around all galaxies for the UHECR-induced cascade model to work with reasonable jet powers. 

\end{abstract}

\maketitle

\section{Introduction}

Anisotropy in the arrival direction distribution of ultrahigh-energy (UHE) cosmic rays (CRs) provides valuable information for understanding the origin of UHECRs, whose sources remain uncertain. The marginal anisotropy reported by the Pierre Auger Observatory (PAO) in the highest energy range \citep[$\gtrsim 6 \times 10^{19}$ eV;][]{Abraham2007Sci318p938,Abreu2010APh34p314,Aab2014arXiv1411.6111} constrains the apparent number density of ultrahigh-energy cosmic-ray (UHECR) sources in local Universe $n_s \gtrsim 10^{-5}-10^{-4}$ Mpc$^{-3}$ in the case of small deflections. It disfavors rare promising source candidates, e.g., Fanaroff-Riley II galaxies and blazars, as a dominant contributor to the observed UHECR intensity \citep{Takami2009Aph30p306,Cuoco2009ApJ702p825,PAO2013JCAP05p009}. Quantitative constraints depend on magnetic deflection and therefore the composition of the UHECRs, because larger deflection angles for heavy nuclei with a given energy provide weaker anisotropy constraints \citep[e.g.,][]{Takami2012APh35p767,PAO2013JCAP05p009}. Furthermore, the dependence of anisotropy on CR energy, if detected, can give statistical evidence for the transient generation of UHECRs \citep{Takami2012ApJ748p9}.

Anisotropy is also useful for examining the composition of UHECRs independently of measurements of maximum atmospheric slant depth $X_{\rm max}$ of extensive air showers induced by UHECRs. If anisotropy is measured for heavy nuclei with energy $E$ and atomic number $Z$, a comparable anisotropy should be produced by protons with energy $E/Z$, because the propagation trajectories of charged particles depend only on particle rigidity when energy loss is negligible \citep{Lemoine2009JCAP11p009}. This statement was tested with $X_{\rm max}$ measurements by the PAO, which has provided the largest UHECR sample available for analysis. The PAO collaboration reports a gradual change of composition to heavy nuclei above the ankle energy for a variety of hadronic interaction models \citep{Abraham2010PRL104p091101,Aab2014arXiv1409.5083P,2014arXiv1409.4809P}. No significant anisotropy is found at lower energies \citep{Abreu2011JCAP06p022}, indicating that either 1) protons dominate the composition at the highest energies and hadronic interaction models should be modified accordingly, 2) the observed anisotropy is a statistical fluctuation, or 3) UHECR sources selectively emit heavy nuclei over a wide energy range.

Gamma rays provide another important probe of UHECR sources.  Electromagnetic cascades are induced by secondary particles produced by interactions of UHECRs with photons of the cosmic microwave background (CMB) and extragalactic background light (EBL) during propagation through intergalactic space. Very-high-energy (VHE; $\gtrsim 100$ GeV) $\gamma$-ray observations have resulted in the detections of several extreme high-frequency-peaked BL Lac objects (EHBLs), most notably 1ES 0229+200, 1ES 1101-232, and 1ES 0347-121, which show very hard VHE spectra  ($\Gamma \lesssim 3;~dN / dE \propto E^{-\Gamma}$) extending to tens of TeV \citep[e.g.,][]{Aharonian2007AA470p475,Aharonian2007AA473L25,Aharonian2007AA475L9}. Moreover, these sources show weak variability, unlike most blazars. VHE $\gamma$ rays can be attenuated by pair-creation interactions with EBL photons while propagating through intergalactic space \citep[][]{Nikishov1962JETP14p393,Gould1966PRL16p252,Stecker1992ApJ390L49}. A possible scenario to overcome the severe absorption that would soften the TeV spectra of these sources is the UHECR-induced cascade model \citep[e.g.,][]{Essey2010APh33p81,Essey2011ApJ731p51,Murase2012ApJ749p63,Razzaque2012ApJ745p196}. In this model, the long ($\sim $ Gpc) energy-loss length from Bethe-Heitler pair creation of UHECR protons with CMB and EBL photons allows the detection of VHE photons with energies above the characteristic EBL attenuation energy at which the optical depth of pair creation with the EBL is unity. This will produce a spectrum that is harder than expected if $\gamma$-rays are emitted at the blazar source. A synchrotron self-Compton model, which is the standard scenario for explaining the spectral energy distribution of BL Lac objects, might still be viable by assuming very large minimum Lorentz factors of electrons \citep[e.g.,][]{Tavecchio2009MNRAS399L59}, though it does not provide a simple explanation for the weak TeV variability of these EHBLs. The spectral hardness above the characteristic EBL attenuation energy, depending on source redshift, can confirm the UHECR-induced cascade model, if identified by next generation $\gamma$-ray telescopes such as Cherenkov Telescope Array \citep{Murase2012ApJ749p63,Takami2013ApJ771L32}.

Interestingly, the UHECR-induced cascade model also requires that UHECRs maintain strong collimation as they travel through intergalactic space. Consequently a strong anisotropy of UHECRs in the directions of EHBLs, if they are powerful UHECR emitters, can be expected in order to produce the observed point-like images and keep the $\gamma$-ray conversion efficiency high \citep[][]{Murase2012ApJ749p63,Razzaque2012ApJ745p196}. The effective extragalactic magnetic field $B_{\rm eff}$ averaged over the UHECR's path is required to be $ \lesssim 10^{-11}$ G (assuming a coherence length $\lambda_{\rm eff} \approx 0.1$ -- 1 Mpc) in order that $10^{19}$ eV proton propagate through intergalactic space from sources at a distance of $\approx 1$ Gpc with deflections $\lesssim 1^\circ$. In fact, much weaker fields of $B_{\rm IGV}\lesssim 10^{-14}$ G in intergalactic voids are required in order that the Bethe-Heitler pairs are not deflected away from the line of sight.

The UHECR-induced cascade model requires, moreover, a huge energy output in UHECRs. An isotropic equivalent CR luminosity $\gtrsim 10^{45}$ erg s$^{-1}$ is needed to reconcile the predicted $\gamma$-ray fluxes with the observed fluxes \citep{Essey2011ApJ731p51,Razzaque2012ApJ745p196,Murase2012ApJ749p63}. This value is much larger than needed for sources to produce the UHECR luminosity at $10^{19}$ eV of \footnote{In this work, the differential CR luminosity $\displaystyle L_{\rm UHECR} = E^2 (d\dot{N}/dE)$ at $10^{19}$ eV is used to follow notation in our previous work \citep{Murase2012ApJ749p63}.}
{\small 
\begin{eqnarray} 
L_{\rm UHECR}^{\rm ave} &=& 3 \times 10^{40} 
\left( \frac{\mathcal{E}(10^{19}~{\rm eV})}{10^{44}~{\rm erg~Mpc}^{-3}~{\rm yr}^{-1}} \right) \left( \frac{n_s}{10^{-4}~{\rm Mpc}^{-3}} \right)^{-1}~~~{\rm erg~s}^{-1}, \nonumber \\
&=& 3 \times 10^{43} \left( \frac{\mathcal{E}(10^{19}~{\rm eV})}{10^{44}~{\rm erg~Mpc}^{-3}~{\rm yr}^{-1}} \right) \left( \frac{n_s}{10^{-7}~{\rm Mpc}^{-3}} \right)^{-1}{\rm erg~s}^{-1},
\label{eq:ave}
\end{eqnarray}}%

{\noindent 
depending on the assumed source density $n_s$. Here $\mathcal{E}(10^{19}~{\rm eV}) \sim 10^{44}$ erg Mpc$^{-3}$ yr$^{-1}$ is the differential CR luminosity density that is required to reproduce the observed flux of $10^{19}$ eV UHECRs \citep{Waxman:1998yy,Berezinsky2006PRD74p043005,Murase2009ApJ690L14,Katz2009JCAP03p020}. The reference values of $n_s = 10^{-4}$ Mpc$^{-3}$ and $10^{-7}$ Mpc$^{-3}$ correspond to the local number density of UHECR sources constrained by anisotropy measurements and that of BL Lac objects \citep{Ajello2014ApJ780p73}, respectively. Thus, small-scale anisotropy is naturally expected in the directions of UHECR sources if they are powerful and rare. 
}

In this paper, we use numerical simulations to provide a general study of the anisotropy in the arrival direction distribution of UHECRs produced by distant powerful UHECR sources. First, we consider  powerful sources of UHECRs located at several representative redshifts and constrain their CR luminosity on the basis of the observed isotropy above $10^{19}$ eV. The effects of magnetic fields associated with cosmic structures are also discussed. Then, we focus on specific objects classified as EHBLs, namely 1ES 0229+200, 1ES 1101-232, and 1ES 0347-121, and derive conditions that the UHECR-induced cascade model must satisfy in order to be consistent with the observed UHECR arrival distribution. We also consider the effects of the Galactic magnetic field (GMF) on the arrival direction distribution of UHECRs.

Throughout this study, we consider only protons as CRs, as motivated by lack of significant anisotropy at $E/Z$ energies \citep{Abreu2011JCAP06p022} in comparison with the marginal ($2$ -- $3 \sigma$) anisotropy claimed at the highest energies \citep{Abraham2007Sci318p938,Abreu2010APh34p314,Aab2014arXiv1411.6111}. It should be also stressed that recent composition measurements also indicate light composition at $\sim 10^{19}$ eV, depending on hadronic interaction models \citep{Aab2014arXiv1409.5083P}. The $\Lambda$CDM cosmology is adopted with the Hubble constant $H_0 = 71$ km s$^{-1}$ Mpc$^{-1}$, matter density normalized by the critical density of $\Omega_{\rm M} = 0.3$, and the cosmological constant in the unit of the critical density $\Omega_{\Lambda} = 0.7$.

This outline of the paper is as follows:  Section \ref{sec:methods} presents the methods to calculate and analyze the arrival direction distribution of UHECRs from a strong UHECR emitter. In Section \ref{sec:ul}, we derive upper limits on the CR luminosity of a strong UHECR source located at various redshifts. Anisotropy in the UHECR sky produced by several representative EHBLs in the UHECR-induced cascade model, and the conditions required to reconcile the cascade model with the observed isotropic sky, are presented in Section \ref{sec:app_ehbls}. The results are discussed in Section \ref{sec:discussion}, and we summarize in Section \ref{sec:summary}.

\section{Calculation and Analysis Methods} \label{sec:methods}

UHECR events in our simulations consist of events from a powerful source as well as events from an isotropic UHECR background. The former events are numerically calculated by considering interactions with cosmic background photons and the deflections of UHECR trajectories by cosmic magnetic fields. The model of cosmic magnetic fields adopted in these simulations is explained in Section \ref{sec:mag}, and the method to calculate the arrival distribution of UHECRs is described in Section \ref{sec:ad}. The number of events from a given source can be calculated by multiplying the CR luminosity of the source, the probability distribution of arrival events, and the exposure geometry of a specific UHECR detector, as described in Section \ref{sec:aperture}. Background events are calculated as events isotropically distributed according to a detection probability that follows the aperture geometry, and whose number is the total number of expected events calculated from the UHECR spectrum under an assumed total exposure minus the number of events from the source. In this study, the analytical fitting formula to the latest PAO spectrum is applied \citep{Auger2013arXiv1307.5059}.

The arrival direction distribution of UHECRs obtained from the simulations is quantified by the cumulative auto-correlation function of the events. We then compare this distribution with an isotropic distribution using the statistical method described in Section \ref{sec:stat}.

\subsection{Cosmic Magnetic Fields} \label{sec:mag}

We classify cosmic magnetic fields into four components defined in terms of UHECR propagation. First, all the UHECRs arriving at the earth are affected by the GMF. The shape of the GMF follows the spiral structure of the Milky Way, and therefore the deflection angles of UHECRs depend on their arrival directions. We consider the GMF only when we focus on specific sources with known positions (Section \ref{sec:app_ehbls}). In this study, a bisymmetric spiral field model with the parametrization of \citet{AlvarezMuniz:2001vf} is adopted for demonstration, although more sophisticated GMF models due to the progress of observations have been recently proposed \citep[e.g.,][]{Pshirkov2011ApJ738p192,Jansson2012ApJ757p14}. Dependence on GMF configurations to the arrival distribution of UHECRs are discussed in detail in \citet{Takami2008ApJ681p1279} and \citet{Takami2010ApJ724p1456}.

Among extragalactic magnetic fields (EGMFs), intergalactic magnetic fields in voids (under-dense regions away from the sources) are poorly known, but lower limits can be placed by searches for $\gamma$-ray pair halos \citep{Neronov2007JETPLett85p579} and pair echoes \citep{Murase2008ApJ686L67}, where one can determine magnetic fields that are strong enough for e$^+$-e$^-$ pairs made by EBL absorption of TeV $\gamma$ rays to be deflected, in order that CMB photons Compton-scattered to GeV energies do not overproduce the Fermi-LAT observations. These void fields $B_{\rm IGV} \gtrsim 10^{-15}$ G for persistent sources, and $B_{\rm IGV} \gtrsim 10^{-18}$ G for sources operating on timescales of the TeV observations are derived from spectral observations  \citep{Neronov2010Science328p73,Tavecchio2011MNRAS414p3566,Dolag2011ApJ727L4,Dermer2011ApJ733L21,Takahashi2012ApJ744L7}. The UHECR-induced cascade model requires $B_{\rm IGV} \lesssim 10^{-14}$ G for reasons noted above, and the effective magnetic field, $B_{\rm eff}$, averaged over structures such as filaments and clusters with small volume-filling fraction and the voids that dominate the volume of the universe 
is $\lesssim 10^{-11}$ G (see \citet{Takami2012ApJ748p9}).

Cosmic structure formation theory indicates that galaxies and other astrophysical objects that could be UHECR sources are typically embedded in dense concentrations of matter. EGMFs in the over-dense structured regions affect the propagation of UHECRs through and out of the structure, which is important to consider in the UHECR-cascade model \citep{Murase2012ApJ749p63} and for synchrotron pair echoes and halos \citep{Oikonomou2014AA568p110}. As a result, for a given beaming-corrected CR luminosity, the isotropic-equivalent CR luminosity $L_{\rm UHECR,MS}^{\rm iso}$\footnote{The subscript MS means isotropic-equivalent CR luminosity affected by magnetic structures (MSs) such as clusters and filaments.} estimated for CRs is significantly smaller than the intrinsic isotropic-equivalent CR luminosity $L_{\rm UHECR}^{\rm iso}$, especially if UHECRs are emitted from collimated jets \citep{Murase2012ApJ749p63}. The former isotropic-equivalent CR luminosity $L_{\rm UHECR,MS}^{\rm iso}$ is related to CR observables and is the CR luminosity estimated from $\gamma$-ray observations in the UHECR-induced cascade model when the beaming angle of CRs is not altered. The latter one is specific for a source population. These two CR luminosities are the same if there is no magnetic structure surrounding UHECR sources. When one takes into account EGMFs around the source, the intrinsic CR luminosity can be $L_{\rm UHECR}^{\rm iso} \sim 10$ -- $100 L_{\rm UHECR,MS}^{\rm iso}$ for sources with a beaming angle of $\sim 0.1$. Accordingly, the number density of observed beamed sources is smaller than that of UHECR sources. Note that these EGMFs does not affect the angular spread of the images of arrival UHECRs as long as the distance of sources is much greater than the size of magnetic structures.

The Milky Way is contained within large-scale structure of matter, including filamentary structures and the local supercluster of galaxies, which will be accompanied by magnetic field. This local EGMF (LEGMF) surrounding the Milky Way also inevitably affects the distribution of arriving UHECRs and is more relevant to diminish the anisotropy of UHECRs assumed to be propagating rectilinearly in intergalactic space, which is satisfied especially in the UHECR-induced cascade model. In this paper we show that this LEGMF can play a key role in reducing the expected anisotropy in the direction of EHBLs in comparison with the observed isotropy at $10^{19}$ eV, as discussed in Section \ref{sec:app_ehbls}. A similar (but larger) magnetic structure was taken into account to isotropize UHECRs emitted from nearby powerful sources inside this structure in order to overcome the Greisen-Zatsepin-Kuz'min \citep{Greisen1966PRL16p748,Zatsepin1966JETP4L78} cutoff \citep{Blasi1999PRD59p023001,Farrar2000PRL84p3527,Ide2001PASJ53p1153}.

Although being poorly known, the properties of the Milky Way's LEGMF are expected to be similar to those magnetic fields found in filamentary structures or clusters of galaxies.  Although it does not have to hold in general, the Copernican principle, in which all galaxies are surrounded by magnetized structures, can be satisfied, as we discuss in Section 5. Following \citet{Murase2012ApJ749p63}, we model this local magnetic structure surrounding our Galaxy as a spherically magnetized region with 2 Mpc radius, magnetic field strength $B_{\rm LEG}$, and magnetic coherence length $\lambda_{\rm LEG} = 100$ kpc. Note that this spherical structure is a simple approximation of a part of cosmic magnetized structures, not intending a structure specific to the galaxy. The value of $B_{\rm LEG}$ depends on where we reside in the local large scale structure. We consider values for the EGMF of $B_{\rm LEG} = 0$, $1$, $10$, and $100$ nG. Note that values of $B_{\rm LEG} = 10$ nG and $100$ nG are expected to be comparable with magnetic fields in filamentary structures and average magnetic fields in clusters of galaxies obtained in detailed simulations \citep{Ryu2008Science320p909}.

\subsection{Arrival Directions of UHECRs from a Powerful Source} \label{sec:ad}

UHECRs emitted from a source suffer energy losses during propagation through intergalactic space to the Earth. We calculate the propagation processes with the method used by \citet{Murase2012ApJ749p63}. The  energy-loss processes that are treated in the Monte-Carlo forward-tracking calculation for UHE protons include Bethe-Heitler pair creation and photomeson production with the photons of the CMB and EBL, and adiabatic cooling by cosmic expansion. The low-IR EBL model of \citet{Kneiske2004AA413p807} is adopted for the calculation. The generation spectrum of UHECRs is assumed to be described by a power-law function with an exponential cutoff given by the expression
\begin{equation}
\frac{d^2N}{dt_g dE_g}(E_g) \propto E_g^{-s} \exp \left( - \frac{E_g}{E_{\rm c}} \right), 
\end{equation}
where $E_g$ is the injection energy, $E_{\rm c}$ is the spectral cutoff energy, and $s$ is the spectral index. The two parameters $E_{\rm c}$ and $s$ depend on the properties of the UHECR source under consideration. In our study of the anisotropy produced by a powerful CR source, we let $E_{\rm c} = 10^{20}$ eV. Because the leptonic interpretation of the spectral energy distribution of BL Lac objects (other than EHBLs) indicates that the maximum acceleration energy of protons is $\sim 10^{19}$ eV \citep{Murase2012ApJ749p63}, we additionally treat the case of $E_{\rm c} = 10^{19}$ eV. Even in this latter case, a fraction of UHECRs are found at energies $\gtrsim 10^{19}$ eV because the spectrum is not sharply truncated at $10^{19}$ eV. As for the spectral index $s$, we choose a value that reproduces the spectral index of the total CR spectrum in the energy range near $10^{19}$ eV \citep[$s = 2.6$; e.g.,][]{Berezinsky2006PRD74p043005}, as well as a hard-spectrum case with $s = 2.0$. The spectrum of UHECRs at the Earth, $d^2N(E)/dtdE$, is determined following the propagation calculation.

The number of detected protons above energy $E_{\rm th}$ that are emitted from a source located in the direction ($l_0, b_0$) in galactic coordinates is estimated as 
\begin{equation}
N(>E_{\rm th}) = \int_{E_{\rm th}}^{\infty} dE 
\int d\Omega \frac{d^3 N}{dE dt d\Omega}(E, \theta, \phi) 
\omega (\delta), 
\end{equation}
where $\theta = \theta(l_0, b_0)$ and $\phi = \phi(l_0, b_0)$ are components in a polar coordinate centered by the source position, and $d\Omega = d\cos \theta d\phi$. The threshold energy considered is $E_{\rm th} = 10^{19}$ eV throughout this paper because CRs above $10^{19}$ eV are widely accepted to be of extragalactic origin, though the transition energy from Galactic CRs and extragalactic CRs is still debated. The exposure of a specific experiment $\omega(\delta)$ (units of time) is a function of declination $\delta = \delta[\theta(l_0, b_0), \phi(l_0, b_0)]$, and is given in the next subsection. The spectral proton injection rate per solid angle from a source in the direction of earth is represented by $d^3 N / dE dt d\Omega$. Provided that only the local EGMF and/or GMF are taken into account, this distribution is separable into the spectrum of UHECRs after propagation in intergalactic space and the angular distribution around the source direction generated by these magnetic fields. Therefore we can write
\begin{equation}
\frac{d^3 N}{dE dt d\Omega}(E, \theta, \phi) 
= \frac{d^2 N}{dE dt}(E) \frac{d n}{d\Omega}(\theta, \phi), 
\end{equation}
because the energy-loss of UHECRs can be neglected in the local magnetic structure surrounding the Milky Way where the propagation path is much shorter than the Bethe-Heitler energy-loss length. In comparison, most of the Bethe-Heitler pair production and photomeson production takes place during transit through intergalactic space.

The function $dn/d\Omega$ is the convolution of the angular distribution of UHECRs generated by magnetic deflections in the LEGMF and the GMF with the angular determination accuracy of UHECR detectors. The angular spreading distribution of UHECRs in the local magnetic field is calculated by the method used in \citet{Murase2012ApJ749p63}, which is essentially the same method as in \citet{Yoshiguchi2003ApJ586p1211} and \citet{Takami2008JCAP06p031}, though energy-loss processes are neglected in the LEGMF and GMF. We inject UHECRs from the origin of coordinates, calculate their trajectories in the magnetized sphere, and record their velocity directions when they reach the boundary of the sphere 2 Mpc away from the center. The deflection angles of UHECRs can be estimated from the difference between their injection directions and the velocity directions. The distribution of the deflection angles can be approximated by a two-dimensional Gaussian function if the deflections are small enough, but this approximation becomes poor when $B_{\rm LEG} \gtrsim 10$ nG, and the distribution then has to be calculated numerically.

The modification of the arrival directions of UHECRs caused by the GMF is calculated by the backtracking method of UHECR propagation in the Milky Way \citep[e.g.,][]{Stanev1996ApJ479p290,Yoshiguchi2003ApJ596p1044}. In this method, UHECRs with the opposite electric charge to protons are injected from the Earth isotropically. The calculation of their propagation is stopped at the boundary of the Milky Way, which is chosen as 40 kpc from the Galactic center, and their velocity directions are recorded. These recorded directions can be regarded as the arrival directions of UHECRs before GMF modification. In order to accurately calculate the modification function for the GMF, we inject $2\times 10^6$ simulated UHECRs in each energy bin with the width of $\Delta \log_{10} E = 0.1$.

The angular determination uncertainty is simulated by a two dimensional Gaussian distribution with a $1^\circ$ 68\% containment radius. This is comparable with the angular arrival uncertainties of UHECRs measured with the PAO \citep{Abraham2007Sci318p938,Abraham2007Aph29p188}.

\subsection{Aperture geometry of UHECR experiments} \label{sec:aperture}

The actual exposure of a ground array is nonuniform in different directions of the sky because the array is at a fixed location on the ground. Since the variation in right ascension in a day can be neglected when discussing anisotropy uncertainties at the $\gtrsim 1\%$ level \citep[e.g.,][]{Abraham2007Aph29p188}, the geometry of the exposure over a much longer time than day scale can simply be estimated as
\begin{equation}
\omega (\delta) \propto \cos (a_0) \cos (\delta) \sin (\alpha_m) + \alpha_m \sin (a_0) \sin (\delta), 
\end{equation}
where 
\begin{eqnarray}
\alpha_m = \left\{
\begin{array}{ll}
0 & {\rm if~\xi~>~1} \\
\pi & {\rm if~\xi~<~-1} \\
\cos^{-1}(\xi)~~ & {\rm otherwise}, 
\end{array}
\right.
\end{eqnarray}
and 
\begin{equation}
\xi = \frac{\cos (\theta_{\rm cut}) - \sin (a_0) \sin (\delta)}
{\cos (a_0) \cos (\delta)}. 
\end{equation}
This differential exposure depends only on the declination of arrival CRs. Here, $a_0$ and $\theta_{\rm cut}$ are the terrestrial latitude of a ground array and the zenith angle for an experimental cut, respectively \citep{Sommers2000Aph14p271}. We adopt the configuration of PAO, $a_0 = -35.2^{\circ}$ and $\theta_{\rm cut} = 60^{\circ}$ \citep{Abreu2010APh34p314}\footnote{The latest PAO data consist of vertical ($\theta_{\rm cut} = 60^{\circ}$) and inclined ($60^{\circ} < \theta < 80^{\circ}$) events \citep{Aab2014arXiv1411.6111}. Nevertheless, this paper considers only vertical events to simulate the PAO in 2013.}, to simulate anisotropy produced by EHBLs in Section \ref{sec:app_ehbls}, because all three EHBLs considered in this study are located in the aperture of the PAO. On the other hand, for a detector-independent study, we consider a uniform exposure to constrain the CR luminosity of a strong source in Section \ref{sec:ul}, where the total exposure can be treated as a parameter.

\subsection{Statistical Quantity} \label{sec:stat}

In this study, anisotropy in the arrival-direction distribution of UHECRs is quantified by calculating an angular auto-correlation function. The angular event distribution from the direction of a specific source is a precise statistical quantity to measure anisotropy produced by powerful sources, as used in \citet{Abreu2010APh34p314} for Centaurus A. This quantity is more sensitive to anisotropy than an angular auto-correlation function and therefore allows one to obtain stronger upper limits of CR luminosity, if source position is known {\it a priori} and the clustering center of UHECRs from a given source is conserved during propagation through the LEGMF. In fact, however, strong event clustering can appear away from the parent source position because the GMF strongly modifies the arrival directions of UHECRs. Indeed, the deflection of focused CRs ($\sim 10^{19}$ eV) by the GMF can be $\sim 30^{\circ}$ \citep[e.g.,][]{Yoshiguchi:2004kd}. In this case, it may be difficult to trace back the source location from the arrival directions. We therefore adopt an angular auto-correlation function that allows us to study anisotropy in the sky without prior knowledge of the clustering center.

The cumulative angular auto-correlation function is defined as 
\begin{equation}
w(< \theta) = \frac{CC(< \theta) - 2 CC'(< \theta) + C'C'(< \theta)}
{C'C'(< \theta)}, 
\end{equation}
where $C$ and $C'$ symbolize simulated CR events and events randomly distributed following the geometry of an experimental exposure $\omega(\delta)$, respectively \citep{Takami2011PThPh126p1123}. $CC(< \theta)$ is the normalized number of (self-)pairs of $C$ within the angular distance of $\theta$. $C'C'(< \theta)$ are also defined in a similar way. $CC'(< \theta)$ is the normalized number of pairs of $C$ and $C'$ within the angular distance of $\theta$. $CC'(< \theta)$ and $C'C'(< \theta)$ correct the inhomogeneous exposure of a dedicated CR experiment and allow us to interpret a cumulative auto-correlation function $w(< \theta)$ as follows: positive correlation for $w(< \theta) > 0$, no correlation for $w(< \theta) = 0$, and negative correlation for $w(< \theta) < 0$. Also, this quantity allows us to directly compare anisotropic signals of UHECRs in different experiments. Since the angular determination accuracy of typical UHECR experiments is $\sim 1^{\circ}$, the  auto-correlation function is evaluated in a grid of at least 180 points with $\Delta \theta = 1^{\circ}$.

\begin{figure*}
\includegraphics[clip,width=0.48\linewidth]{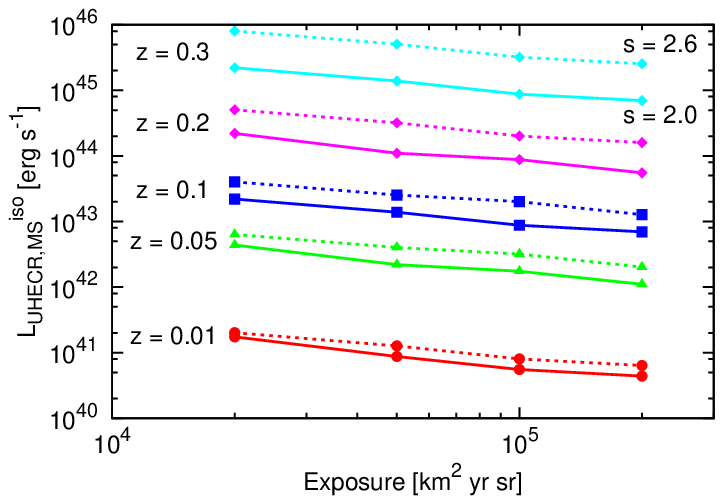}
\includegraphics[clip,width=0.48\linewidth]{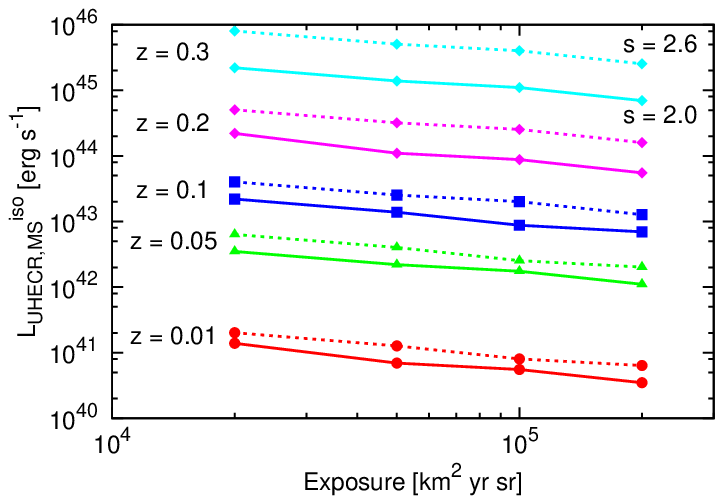} \\
\includegraphics[clip,width=0.48\linewidth]{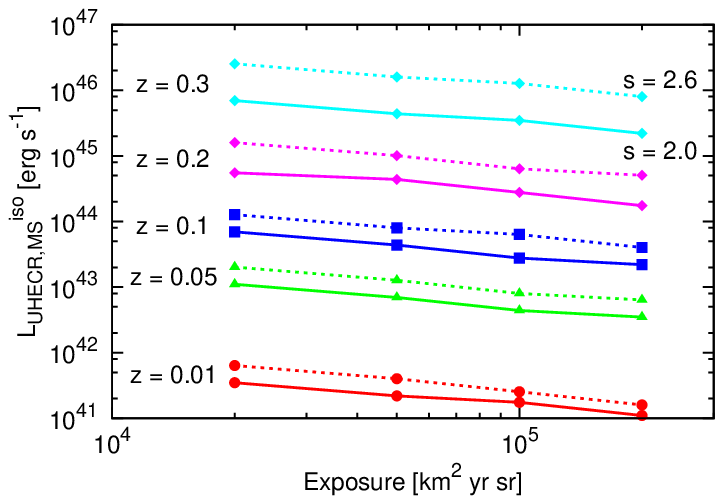}
\includegraphics[clip,width=0.48\linewidth]{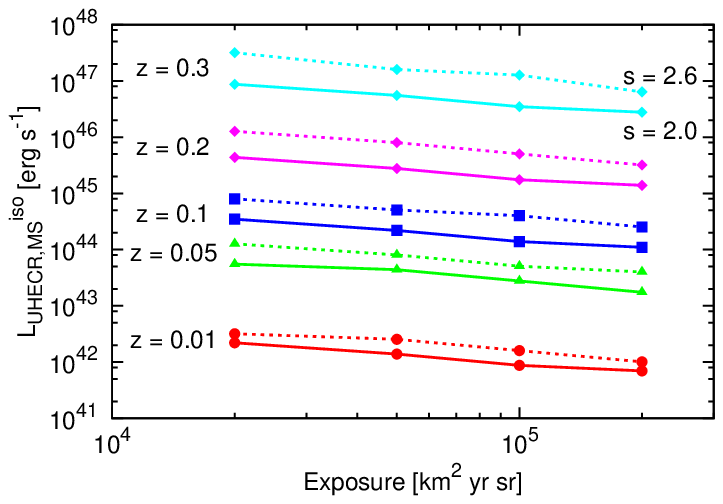}
\caption{The 95\% CL upper limits of the UHECR proton luminosity $L_{\rm UHECR,MS}^{\rm iso} = E^2 dN / dE$ at $10^{19}$ eV as a function of the exposure of a UHECR experiment with uniform aperture. The cases of sources located at five representative redshifts and for two spectral indices ($s = 2.0$ [{\it solid lines}] and $s = 2.6$ [{\it dotted lines}]) are shown. The assumed strengths of the LEGMF are $B_{\rm LEG} = 0$ nG ({\it upper left}), $1$ nG ({\it upper right}), $10$ nG ({\it lower left}), and $100$ nG ({\it lower right}). The cutoff energy of injected UHECR spectra is $E_{\rm c} = 10^{20}$ eV. The exposure reported in 2013 for the PAO is $\sim 3.2 \times 10^4$ km$^2$ sr yr \citep{Auger2013arXiv1307.5059}.}
\label{fig:Llim20}
\end{figure*}

\begin{figure*}
\includegraphics[clip,width=0.48\linewidth]{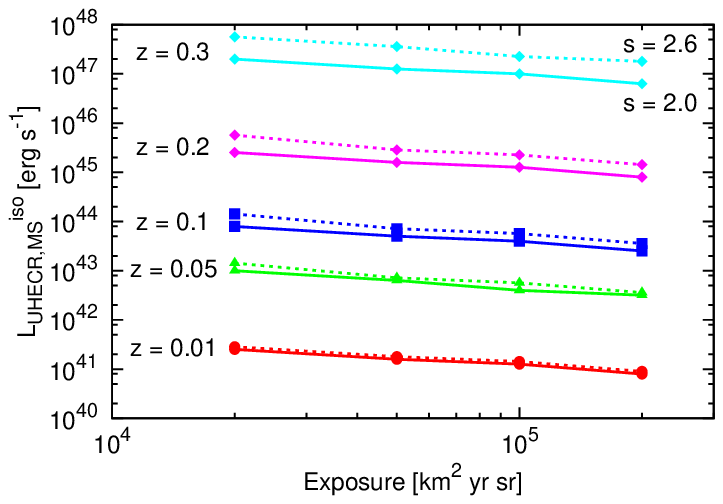}
\includegraphics[clip,width=0.48\linewidth]{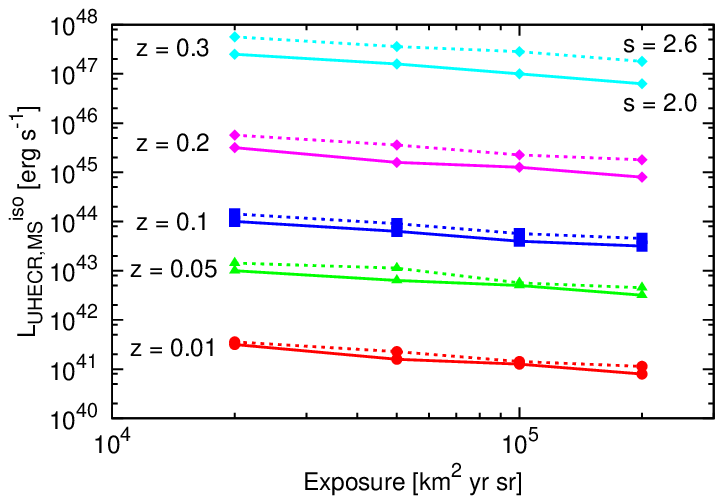} \\
\includegraphics[clip,width=0.48\linewidth]{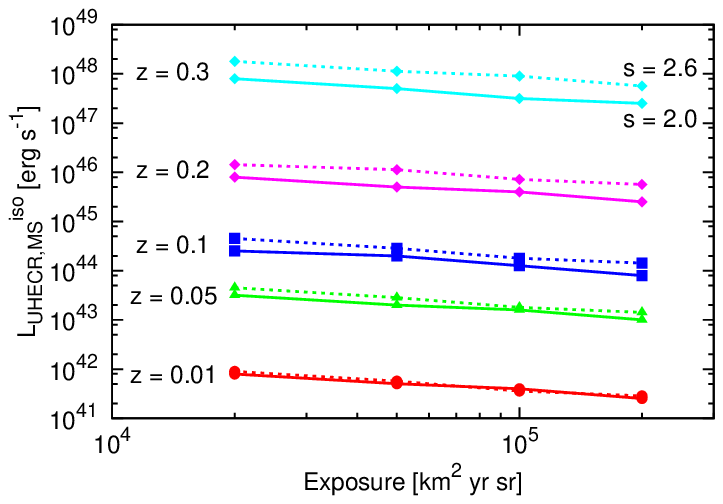}
\includegraphics[clip,width=0.48\linewidth]{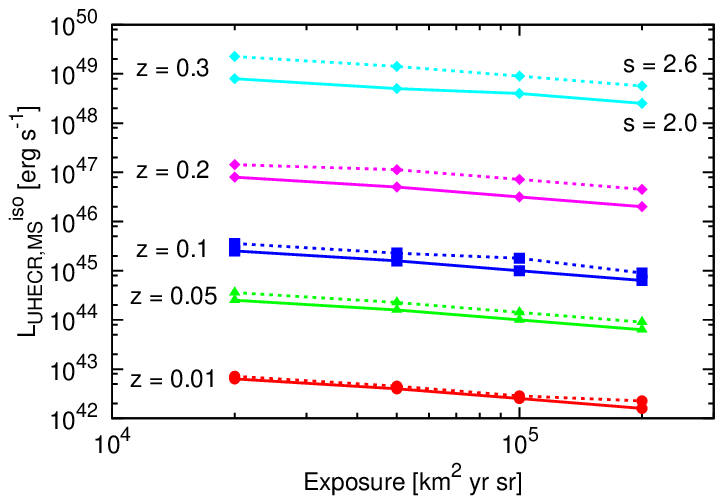}
\caption{Same as Figure \ref{fig:Llim20}, but for $E_{\rm c} = 10^{19}$ eV.}
\label{fig:Llim19}
\end{figure*}

\section{Upper limits on $10^{19}$ eV cosmic-ray luminosity} \label{sec:ul}

\subsection{Results}

The observed isotropy of UHECRs at $\approx 10^{19}$ eV can constrain the UHECR source luminosity $L_{\rm UHECR,MS}^{\rm iso} = E^2 dN/dE$, which is the differential luminosity of CRs at $10^{19}$ eV. The early results of the PAO showed that the arrival distribution of UHECRs above $10^{19}$ eV is consistent with an isotropic distribution at 95\% confidence level (CL) by using pair counts, which is equivalent to an angular auto-correlation function \citep{Mollerach2008ICRC30v4p279}.

The upper limits on the CR luminosity for a source of UHECRs with spectral index $s$ that propagate through a given LEGMF with strength $B_{\rm LEG}$, as a function of the total exposure of a hypothetical experiment with uniform aperture $T$, are estimated as follows: We simulate 1000 sets of arriving UHECR and calculate the corresponding cumulative auto-correlation functions $w(< \theta)$ for a given $L_{\rm UHECR,MS}^{\rm iso}$. These calculations provide a distribution of $w(< \theta)$ for a given $\theta$. We define the upper limit of the CR luminosity of the source for a given $\theta$ as the maximum CR luminosity for which the area of the distribution with $w(< \theta) \geq 0$ is less than 95\% after surveying $w(< \theta)$ as a function of $L_{\rm CR,MS}^{\rm iso}$ in steps of $\Delta \log_{10} L_{\rm UHECR}^{\rm iso} = 0.1$. This upper limit depends on a focused $\theta$. Motivated by the isotropy observed on all angular scales, we define the upper bound of the CR luminosity of the source as the maximum CR luminosity for which the area of the distribution with $w(< \theta) \geq 0$ is less than 95\% at all angular scales. Throughout this paper, we call this the 95\% CL upper bound of CR luminosity.

We consider hypothetical sources placed at five specific redshifts, namely $z = 0.01$, $0.05$, $0.1$, $0.2$, and $0.3$. The source with $z = 0.01$, at a distance of $\approx 40$ Mpc,  is located within the Greisen-Zatsepin-Kuz'min (GZK) radius of $\sim 100$ Mpc \citep[][]{Greisen1966PRL16p748,Zatsepin1966JETP4L78}. The case of $z=0.05$ corresponds to the maximum source distance for which UHECR protons with measured energies of $6 \times 10^{19}$ eV could originate \citep[e.g.,][]{Takami2012ApJ748p9}. The redshift $0.2$ case is comparable with the values of the specific EHBLs considered in this study, and the Bethe-Heitler pair-production energy-loss time for $E_p\gtrsim 10^{18}$ eV protons corresponds to $\lesssim H_0^{-1}$ at $z \approx 0.3$.

Figure \ref{fig:Llim20} shows the upper limits of the CR luminosity of a source in the case $E_c = 10^{20}$ eV as a function of total exposure of a hypothetical experiment with a uniform aperture. The four panels show the cases of the LEGMF strength $B_{\rm LEG} = 0$ nG ({\it upper left}), $1$ nG ({\it upper right}), $10$ nG ({\it lower left}), and $100$ nG ({\it lower right}), and are plotted for two representative spectral indices $s = 2.0$ ({\it solid lines}) and $2.6$ ({\it dotted lines}). Note that the exposure of the PAO is $\sim 3.2 \times 10^4$ km$^2$ sr yr at 2013 \citep{Auger2013arXiv1307.5059}\footnote{The latest exposure of the PAO is $51753$ km$^2$ sr yr and $14699$ km$^2$ sr yr for vertical and inclined events, respectively \citep{Aab2014arXiv1411.6111}, which was posted on arXiv while this manuscript was being prepared. Throughout this paper, we simulate the PAO in 2013.}, that of the TA is $\sim 3.1 \times 10^3$ km$^2$ sr yr \citep{Abu-Zayyad2013ApJ777p88}, and the exposure that will be achieved by the JEM-EUSO mission for three years in space is estimated to be $\sim 2 \times 10^5$ km$^2$ sr yr  \citep{Adams2013APh44p76}.

The upper limits on CR luminosity scale inversely with the square root of total exposure as long as the number of events originating from a focused powerful source is small enough compared to the total number of events. The significance of excess in the number of events over isotropic background can be estimated from the ratio of the number of excess events and the square root of background events within a radius comparable with the typical deflection angles of UHECRs. The numbers of both events are simply proportional to the total exposure of an experiment. Since the upper limit of CR luminosity is just below the luminosity required to realize significant anisotropy, we can apply the upper limits in the discussion of UHECR sources.

The calculated upper limits depend on the spectral index of UHECRs made by a focused powerful source. A steep-spectrum case leads to a larger upper limit since there are fewer higher energy UHECRs, which enhance the anisotropy compared to an equal number of lower-energy CRs. A higher luminosity is therefore required for a soft source to provide the same anisotropy as for a hard source.

The LEGMF with $B_{\rm LEG} = 1$ nG does not affect the results compared to the case of no magnetic field around the Milky Way, because the deflection angles of UHECR protons with $\sim 10^{19}$ eV are comparable with the uncertainty of determining the arrival directions of UHECRs by current UHECR detectors. Above this value, the local EGMF smears out an event cluster produced by a powerful source. As a result, the upper limit of CR luminosity to achieve the observed isotropy becomes higher, and is about 3  and 20 times higher for $B_{\rm LEG} = 10$ nG and $B_{\rm LEG} = 100$ nG, respectively, than the upper limits for the case of no local magnetic field.

The CR luminosity upper limit of a source in the local universe can be estimated from simple interpolation by using the approximation that the comoving/luminosity distance within $z = 0.05$ can be well approximated by $c z / H_0$ within a few percent level. Consequently, the CR luminosity limit is proportional to $z^2$ in the local universe.

The upper limits of the CR luminosity become higher in the cases of $E_{\rm c} = 10^{19}$ eV, as shown in Figure \ref{fig:Llim19}. This is because the fraction of the luminosity of CRs to contribute to the observed CRs above $10^{19}$ eV is smaller due to the small $E_{\rm c}$ than that for $E_{\rm c} = 10^{20}$ eV. The difference between an upper limit value for $E_{\rm c} = 10^{19}$ eV and that for $E_{\rm c} = 10^{20}$ eV becomes large if the redshift of a source approaches $z = 0.3$, which is comparable with the energy-loss length of the Bethe-Heitler pair-creation process for protons with $E_p\approx 10^{19}$ eV.

Here, the effects of EGMFs other than the LEGMF on our general upper limits are mentioned. First, EGMFs surrounding sources can increase the constrained $L_{\rm UHECR,MS}^{\rm iso}$ compared to $L_{\rm UHECR}^{\rm iso}$. Although the upper limits of $L_{\rm UHECR,MS}^{\rm iso}$ remain unchanged for the isotropic sources of UHECRs, upper limits on the intrinsic isotropic-equivalent CR luminosity may be 10 -- 100 times larger than those of $L_{\rm UHECR,MS}^{\rm iso}$, depending on the opening angle of jets and the configuration of the magnetic structure if CRs originate from beaming sources such as blazars \citep{Murase2012ApJ749p63}. Higher CR luminosities are more challenging for the UHECR-induced cascade model to work.

\subsection{Implications}

The simple scaling of the CR luminosity constraints can rule out the existence of typical persistent UHECR sources in the very local universe. The observed isotropy is consistent with the existence of UHECR sources with average CR luminosity $L_{\rm UHECR}^{\rm ave}$ within the distance where the CR luminosity upper limit $L_{\rm UHECR,MS}^{\rm iso}$ is larger than the average CR luminosity. This implies that the characteristic CR source distance 
\begin{equation} \small
d \gtrsim 24 \left( \frac{n_s}{10^{-4}~{\rm Mpc}^{-3}} \right)^{-1/2} 
\left( \frac{L_{\rm UHECR,MS}^{\rm iso,ul}(z = 0.01)}{10^{41}~{\rm erg~s}^{-1}} \right)^{-1/2}~~{\rm Mpc}, \label{eq:rlim}
\end{equation}
where the right-hand side is normalized by the source number density estimated from observations of UHECRs above $6 \times 10^{19}$ eV \citep{Takami2009Aph30p306}. Here, $\mathcal{E}(10^{19}~{\rm eV}) = 10^{44}$ erg Mpc$^{-3}$ yr$^{-1}$ is applied and $L_{\rm UHECR,MS}^{\rm iso,ul}(z = 0.01)$ is the CR luminosity upper limit of a source located at $z = 0.01$ at $10^{19}$ eV, which is used for interpolation to lower and higher redshift.

The upper limits of CR luminosity allow us to estimate the source number density of UHECRs with energy $E_p\approx 10^{19}$ eV, 
if $n_s$ is regarded as a free parameter. 
First, equation \ref{eq:rlim} means that one should detect anisotropy from the nearest source 
in the cumulative auto-correlation function, if its distance is smaller than the critical distance 
$d_{\rm crit}$ defined as the right-hand side of equation \ref{eq:rlim}. Note that, even if all the sources 
have the same luminosity $L_{\rm UHECR, MS}^{\rm iso}$ and most of the UHECRs 
originates from unresolved distant sources, the anisotropy should still  be seen if their exists 
a very nearby UHECR source thanks to large statistics at this energy. Then, the number of 
nearby UHECR sources within the distance $d_{\rm crit}$ is given 
by $(\Omega / 3) d_{\rm crit}^3 n_s$, where $\Omega \approx 2 \pi$ is the field-of-view of the PAO. 
The absence of a UHECR source within the critical distance $d_{\rm crit}$ 
is represented as $(\Omega / 3) d_{\rm crit}^3 n_s \lesssim 1$. 
This inequality indicates 
\begin{equation}
n_s \gtrsim 8 \times 10^{-4} \left( \frac{\Omega}{2 \pi} \right)^2  
\left( \frac{L_{\rm UHECR,MS}^{\rm iso,ul}(z = 0.01)}{10^{41}~{\rm erg~s}^{-1}} \right)^{-3}~~{\rm Mpc},  \label{eq:ns19}
\end{equation}
whose numerical factor is much larger than the source number density of 
the highest energy cosmic rays ($> 6 \times 10^{19}$ eV). 
Although this number density is in the very local universe, the source number density averaged 
over larger scale is close to this value because the local overdensity of galaxies is just less than 
a factor of 2 from the results of galaxy surveys \citep{Blanton2001Aph15p275} . 
Importantly, this lower limit of the source number density becomes larger as the upper limit of CR luminosity becomes stronger.

Figure \ref{fig:Llim20} shows $L_{\rm UHECR,MS}^{\rm iso,ul}(z = 0.01) \sim 10^{41}$ erg s$^{-1}$ is already achieved by the PAO for $B_{\rm LEG} \lesssim 1$ nG. Then, interestingly, the numerical value in equation (\ref{eq:ns19}) is larger than that of radio galaxies (Fanaroff-Riley I $+$ II), $\sim 10^{-4}$ Mpc$^{-3}$ \citep{Padovani1990ApJ356p75,vanVelzen2012AA544p18}. This numerical value is also consistent with an early estimation of the source number density of UHECRs with $\sim 10^{19}$ eV with a structured extragalactic magnetic field model \citep{Takami2009Aph30p306}. Once $L_{\rm UHECR,MS}^{\rm iso,ul}(z = 0.01) \sim 10^{41}$ erg s$^{-1}$ is confirmed through CR experiments, we can find that UHECRs with the energy of $10^{19}$ eV should be produced in more common sources such as normal galaxies, if the UHECR sources are steady.

Actually, these constraints depend on the upper limit of CR luminosity, which essentially requires the understanding of the LEGMF. Thus, the observations of the LEGMF via Faraday rotation measurements by Square Kilometer Array\footnote{http://www.skatelescope.org} should also be relevant. The upper limit value of $L_{\rm UHECR,MS}^{\rm iso,ul} \sim 10^{41}$ erg s$^{-1}$ is achievable in the near future for $B_{\rm LEG} \lesssim 10$ nG even in the cases of $E_c = 10^{19}$ eV. Equation (\ref{eq:rlim}) indicates that UHECR source candidates with a typical luminosity needed to power the UHECRs are unlikely to be found within $\sim 20$ Mpc when $B_{\rm LEG} \lesssim 10$ nG, if UHECR emission is steady and all the sources of UHECRs with $\sim 10^{19}$ eV  are those of the highest energy CRs, i.e., have the same number density $\sim 10^{-4}$ Mpc$^{-3}$. Therefore, even in the strong magnetic-field case, nearby source candidates are possible to be ruled out, showing the power of the isotropy constraints.

If $B_{\rm LEG}\approx 100$ nG, Figs. 1 and 2 show that $L_{\rm UHECR,MS}^{\rm iso}$ at $z = 0.01$ reaches $\approx 2 \times 10^{42}$ erg s$^{-1}$ in the near future. In this case, the corresponding source density $n_s \gtrsim 10^{-7}$ Mpc$^{-3}$, 
and becomes consistent with the source number density indicated from $\gtrsim 6 \times 10^{19}$ eV data and that of Fanaroff-Riley I galaxies (this is still inconsistent with on-axis objects such as blazars).

The existence of powerful UHECR emitters has been motivated by EHBLs.  When a BL Lac blazar points in our direction and UHECRs escape from the structured regions in a highly collimated beam, we see an unusual source like 1ES 0229+200, 1ES 0347-121, or 1ES 1101-232. The local blazar density is $n_s \sim 10^{-7}$ Mpc$^{-3}$ \citep{Ajello2014ApJ780p73}, indicating no blazars within the GZK radius. Equations (\ref{eq:rlim}) and (\ref{eq:ns19}) imply that blazars as beamed UHECR sources with $L_{\rm UHECR}^{\rm ave}\sim3\times10^{43}$ erg s$^{-1}$ (see equation \ref{eq:ave}) are also ruled out by our isotropy constraints assuming $B_{\rm LEG} \lesssim 10$ nG. However, a large fraction of BL Lac objects, even though at a similar redshift range, may not have the UHECR cascade effects because CRs can be completely deflected and isotropized by magnetic fields around the source. Then, the vast majority of radio galaxies with jets become unbeamed, quasi-isotropic sources of CRs.  These are essentially misaligned blazars according to the unification scenario \citep{Urry1995PASP107p803}, with a factor $\sim 10^2$ -- $10^3$ more radio sources than for every beamed $\gamma$-ray counterpart with $n_s \sim 10^{-7}$ Mpc$^{-3}$ in the local universe. Many of these misaligned blazars, like Centaurus A, may have their escaping UHECRs isotropized. In this case, the isotropy limits can be applied straightforwardly. Since the source density of radio galaxies is $n_s \sim 10^{-4}$ Mpc$^{-3}$, even the possibility of misaligned blazars may be disfavored by our isotropy limits, depending on the strength of the LEGMFs. Note that anisotropy is reported for Centaurus A at the highest energies, but not at $10^{19}$ eV ranges. The isotropy constraint achievable at present or in the near future should be important to test if steady sources such as Centaurus A are UHECR sources, as pointed out in \citet{Takami2009Aph30p306}.

Following the requirement of the point-like $\gamma$-ray images of EHBLs in the UHECR-induced cascade model, intergalactic magnetic fields between magnetized regions around EHBLs and the Milky Way have been neglected. However, since the UHECR-induced cascade model is just a motivation to consider powerful UHECR emitters, we can instead consider cases where these magnetic fields can significantly affect the propagation of UHECRs emitted from a powerful source. Imagine that UHECRs are emitted in a cone with a certain opening angle $\theta_{\rm op}$. If the deflection of UHECR trajectories in intergalactic space is smaller than the opening angle, the amount of UHECRs penetrating the LEGMF is unchanged, but intervening magnetic fields give deflections in addition to LEGMFs in structured regions. 
On the other hand, if the deflection is larger than the opening angle, that is, 
\begin{equation}
B_{\rm eff} \lambda_{\rm eff}^{1/2} > 5 \left( \frac{E_p}{10^{19}~{\rm eV}} \right) 
\left( \frac{d}{100~{\rm Mpc}} \right)^{-1/2} 
\left( \frac{\theta_{\rm op}}{0.2} \right) ~~~{\rm nG~Mpc}^{1/2}, 
\label{eq:beff}
\end{equation}
where $d$ is the distance of a source, some of UHECRs escape from the cone during propagation in intergalactic space. Intergalactic magnetic fields provide not only an additional deflection but also reduce the flux of UHECRs entering the Milky Way. In this case more sophisticated numerical calculations of UHECR propagation in intergalactic space are needed to estimate the amount of the escaping CRs and then to constrain the CR luminosity.

In order that intergalactic magnetic fields do not affect the critical distance and the source number density of UHECRs with $\sim 10^{19}$ eV estimated in equations (\ref{eq:rlim}) and (\ref{eq:ns19}), the effective intergalactic magnetic fields should be sufficiently weak. The condition is that UHECRs emitted from the nearest source in a cone with the opening angle $\theta_{\rm op}$ reach the Milky Way without losses. Following the discussion deriving equations (\ref{eq:rlim}) and (\ref{eq:beff}), the required property of effective magnetic fields is 
{\small 
\begin{eqnarray}
B_{\rm eff} \lambda_{\rm eff}^{1/2} &\lesssim& 0.93 
\left( \frac{E_p}{10^{19}~{\rm eV}} \right) 
\left( \frac{n_s}{10^{-4}~{\rm Mpc}^{-3}} \right)^{1/4} 
\left( \frac{\mathcal{E}(10^{19}~{\rm eV})}{10^{44}~{\rm erg~Mpc}^{-3}~{\rm yr^{-1}}} \right)^{-1/4} 
\nonumber \\
&& ~~~\times 
\left( \frac{L_{\rm UHECR,MS}^{\rm iso,ul}(z = 0.01)}{10^{41}~{\rm erg}~{\rm s}^{-1}} \right)^{1/4} \left( \frac{\theta_{\rm op}}{0.2} \right) ~~{\rm nG~Mpc}^{1/2}. 
\end{eqnarray}}%
{\noindent 
Here, $n_s = 10^{-4}$ Mpc$^{-3}$ is used for conservative estimation. Given that Faraday rotation measurements suggest $B_{\rm eff} \lambda_{\rm eff}^{-1/2} \lesssim 10~{\rm nG~Mpc}^{1/2}$ \citep{Kronberg1994RepProgPhys57p325}, the source number density of UHECRs with $\sim 10^{19}$ eV can be larger than that of UHECRs at the highest energies. 
}

The upper limits of the CR luminosity of local UHECR sources constrain the energy conversion rate to UHECRs for a powerful source. If UHECRs are accelerated up to the maximum energy of $E_{\rm max}$ in a relativistic jet with the bulk Lorentz factor of $\Gamma$, the isotropic-equivalent total luminosity of the jet $L_{\rm tot}^{\rm iso}$ and corresponding magnetic luminosity $L_{B}^{\rm iso}$ should satisfy $L_{\rm tot}^{\rm iso} > L_B^{\rm iso} > 2 \times 10^{45} Z^{-2} \Gamma^2 (E_{\rm max} / 10^{20}~{\rm eV})^2$ erg s$^{-1}$, where $Z$ is the atomic number of CRs \citep{Norman1995ApJ454p60,Waxman2004Prama62p483,Pe'er2009PRD80p123018}. The conversion ratio from jet energy to CR energy is defined as the ratio of $L_{\rm tot}^{\rm iso}$ and the intrinsic isotropic-equivalent CR luminosity $L_{\rm UHECR}^{\rm iso}$. When the EGMF surrounding UHECR sources are taken into account, the intrinsic isotropic-equivalent CR luminosity is parameterized as $L_{\rm UHECR}^{\rm iso} = \xi L_{\rm UHECR,MS}^{\rm iso}$, where $\xi$ ranges from $1$ (no such an EGMF) to $200 (\theta_{\rm op} / 0.2)^{-2}$ (when CRs are perfectly isotropized by EGMFs). For a powerful source at $z = 0.01$, its CR luminosity is maximally $L_{\rm UHECR,MS}^{\rm iso}\lesssim 10^{42.5}$ erg s$^{-1}$ at present even for a strong magnetic field surrounding the Milky Way with $B_{\rm LEG} = 100$ nG. This indicates that the energy conversion rate to UHECRs can be as high as $L_{\rm UHECR}^{\rm iso} / L_{\rm tot}^{\rm iso} \lesssim 2 \times 10^{-4} Z^2 (\Gamma / 10)^{-2} (E_{\rm max} / 10^{20}~{\rm eV})^{-2} (\xi / 10)$. For reference, the energy conversion rate of an average UHECR source is $L_{\rm CR}^{\rm iso} / L_{\rm tot}^{\rm iso} < 2 \times 10^{-6} Z^2 (\Gamma / 10)^{-2} (E_{\rm max} / 10^{20}~{\rm eV})^{-2} (\xi / 10)$.

\section{Applications to Extreme HBLs} \label{sec:app_ehbls}

\begin{table*}
\centering
\caption{Sample of EHBLs and their parameters}
\begin{tabular}{lccc} \hline \hline
Source Name & 1ES 0229+200 & 1ES 0347-121 & 1ES 1101-232 \\
$z$ & 0.140 & 0.188 & 0.186 \\
$l$ [deg] & $152.97$ & $201.93$ & $273.19$ \\
$b$ [deg] & $-36.61$ & $-45.71$ & $33.08$ \\
$L_{\rm UHECR}^{\rm iso}$ [$10^{44}$ erg s$^{-1}$] ~~($E_c = 10^{20}$ eV, $s = 2.0$) & $4.4$ & $5.7$ & $5.7$ \\
~~~~~~~~~~~~~~~~~~~~~~~~~~~~~~~~~~~~~~($E_c = 10^{20}$ eV, $s = 2.6$) & $4.8$ & $6.4$ & $6.4$ \\
~~~~~~~~~~~~~~~~~~~~~~~~~~~~~~~~~~~~~~($E_c = 10^{19}$ eV, $s = 2.0$) & $6.0$ & $8.0$ & $8.0$ \\
~~~~~~~~~~~~~~~~~~~~~~~~~~~~~~~~~~~~~~($E_c = 10^{19}$ eV, $s = 2.6$) & $3.5$ & $5.0$ & $4.7$ \\
$L_{\rm CR}^{\rm iso}$ [$10^{45}$ erg s$^{-1}$] ~~~~~~~~($E_c = 10^{20}$ eV, $s = 2.0$) & $2.0$ & $2.6$ & $2.6$ \\
~~~~~~~~~~~~~~~~~~~~~~~~~~~~~~~~~~~~~~($E_c = 10^{20}$ eV, $s = 2.6$) & $3.0$ & $4.0$ & $4.0$ \\
~~~~~~~~~~~~~~~~~~~~~~~~~~~~~~~~~~~~~~($E_c = 10^{19}$ eV, $s = 2.0$) & $3.0$ & $4.0$ & $4.0$ \\
~~~~~~~~~~~~~~~~~~~~~~~~~~~~~~~~~~~~~~($E_c = 10^{19}$ eV, $s = 2.6$) & $3.9$ & $5.6$ & $5.2$ \\
\hline \hline
\end{tabular}
\tablecomments{\centering Integrated CR luminosity defined as $\displaystyle L_{\rm CR}^{\rm iso} \equiv \int_{10^{18}~{\rm eV}}^{10^{21}~{\rm eV}} dE_g E_g \frac{d^2N}{dt_gdE_g}$.}
\label{tab:ehbls}
\end{table*}

We have generally derived the constraints on the CR luminosity of a powerful source from the observed isotropy at $\sim 10^{19}$ eV. In this section, we specifically focus on three known EHBLs as powerful UHECR emitters, that is, 1ES 0229+200 \citep{Aharonian2007AA475L9}, 1ES 0347-121 \citep{Aharonian2007AA473L25}, and 1ES 1101-232 \citep{Aharonian2007AA470p475}, all of which have been detected by the High Energy Spectroscopic System (H.E.S.S.). The spectra of these objects in the VHE range can be well reproduced by the UHECR-induced cascade model \citep[e.g.,][]{Essey2011ApJ731p51,Murase2012ApJ749p63}. The isotropic-equivalent CR luminosities to reproduce the observed flux are shown in Table \ref{tab:ehbls} with their redshift and positions. The integrated CR luminosities are typically $\sim 10^{45}$ -- $10^{46}$ erg s$^{-1}$, which are consistent with previous estimation \citep{Essey2011ApJ731p51,Razzaque2012ApJ745p196,Murase2012ApJ749p63}. The expected numbers of UHECR events from EHBLs are analytically estimated in \citet{Razzaque2012ApJ745p196}. In this section, we examine the anisotropy in the arrival direction distribution of UHECRs expected from these objects in the UHECR-induced cascade model with the effects of the LEGMF taken into account. We investigate how such anisotropy can be reconciled with the observed isotropic distribution. Since we focus on specific sources, we can also consider the modifications of the arrival directions of UHECRs by the GMF based on a specific GMF model.

\begin{figure*}
\includegraphics[clip,width=0.48\linewidth]{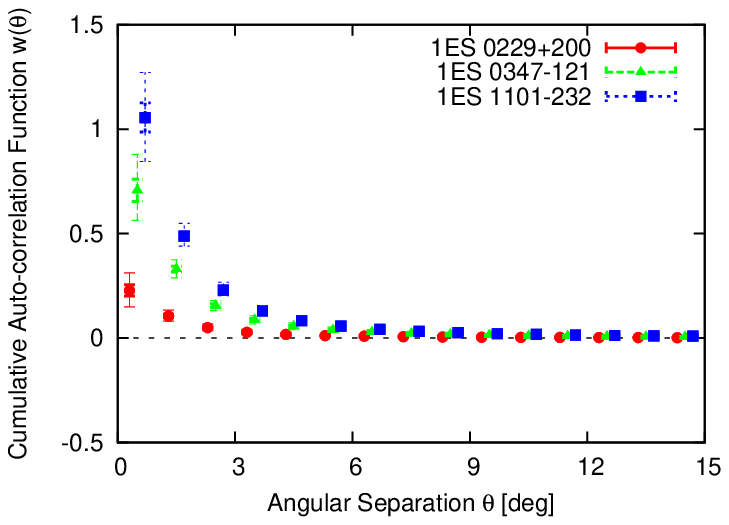}
\includegraphics[clip,width=0.48\linewidth]{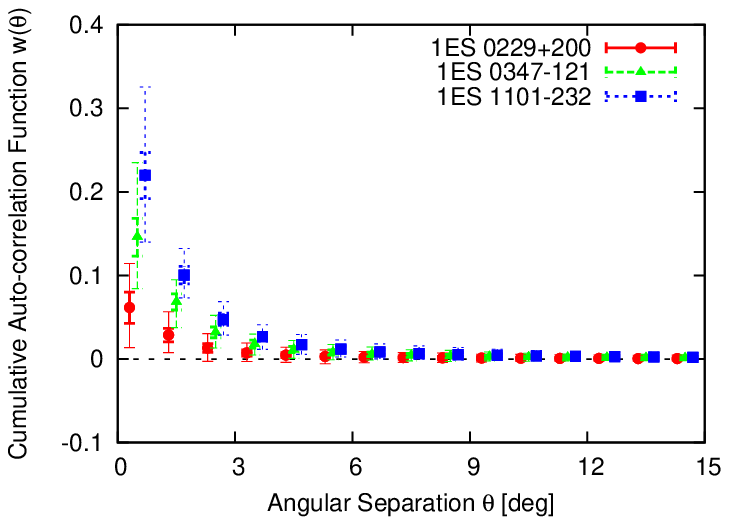}
\caption{Cumulative angular auto-correlation functions of UHECRs consisting of isotropic background UHECRs and a source contribution for the EHBL identified by the legend and defined by the properties given in Table \ref{tab:ehbls}. The two error bars represent $1 \sigma$ ({\it thin}) and $3 \sigma$ ({\it thick}) errors, respectively. The aperture geometry of PAO with the total exposure of $31645$ km$^2$ yr sr to simulate the observational situation of PAO in 2013. The CR spectra are assumed to have $E_{\rm c} = 10^{20}$ eV with $s = 2.0$ ({\it left}) and $2.6$ ({\it right}). Neither the local magnetic field around the Milky Way nor the GMF are taken into account.}
\label{fig:ac20}
\end{figure*}

We infer the anisotropy produced by the EHBLs from the constraints on the CR luminosity for the case of a source located at $z = 0.2$ shown in Figures \ref{fig:Llim20} and \ref{fig:Llim19}. In the case of $E_{\rm c} = 10^{20}$ eV, the upper limits of the CR luminosity for $B_{\rm LEG} \lesssim 10$ nG are comparable with or smaller than the required CR luminosity for $s = 2.0$ (for a total exposure of $T = 31645$ km$^2$ yr sr), and therefore significant anisotropy in the direction of each EHBL is expected. However, if the Milky Way is surrounded by a strongly magnetized medium with $B_{\rm LEG} = 100$ nG, anisotropy in the CR sky is not expected. Even if the CR spectrum is steep, i.e., $s = 2.6$, a significant event cluster appears in the direction of each EHBL in the cases of $B_{\rm LEG} \lesssim 1$ nG. If the EHBLs accelerate protons only up to $10^{19}$ eV, anisotropy may not appear because the required CR luminosity is smaller than the upper limits of CR luminosity at present. However, in the near future, the upper limit can reach the required luminosity if the LEGMF is weak enough.

In order to estimate anisotropy expected in the data taken by a current observatory more accurately, we should consider the aperture geometry of the observatory in the simulations. Figure \ref{fig:ac20} shows the cumulative angular auto-correleation functions calculated from the arrival direction distribution of UHECRs which consist of isotropic background and the contribution of each EHBL, as indicated in the legend. In these simulations the aperture geometry of the PAO with total exposure of $31645$ km$^2$ yr sr is applied to simulate the observational situation of the PAO in 2013 \citep{Auger2013arXiv1307.5059}. The spectrum of UHECRs produced by the sources has the cutoff energy of $E_c = 10^{20}$ eV with $s = 2.0$ ({\it left}) and $s = 2.6$ ({\it right}). Neither the LEGMF nor the GMF are taken into account.

In both cases the auto-correlation functions are inconsistent with zero at small angular scales, and therefore strong anisotropy over the $3\sigma$ excess is predicted at small scale in the directions of these EHBLs. As long as UHECRs around ${10}^{19}$~eV and ${10}^{20}$~eV come from the same source population, rare and powerful sources cannot be typical as the origin of UHECRs. The significance of anisotropy is systematically smaller in the case of the steep spectrum ($s = 2.6$) because the CR luminosity contributing to UHECRs above $10^{19}$ eV is smaller despite  the comparable CR luminosity implied by the gamma-ray data.

The effect of the PAO aperture geometry appears in the difference of the strength of the anisotropic signals of the three EHBLs. The EHBL 1ES 0229+200 is located in the northern terrestrial hemisphere, which is the edge of the PAO aperture. Since exposure is small in this direction, the anisotropic signal produced by this object is small. The other two EHBLs are located in the southern terrestrial hemisphere, and therefore their anisotropic signals are relatively large. The difference between these two signals originate from the different declinations of these objects.

\begin{figure*}
\includegraphics[clip,width=0.48\linewidth]{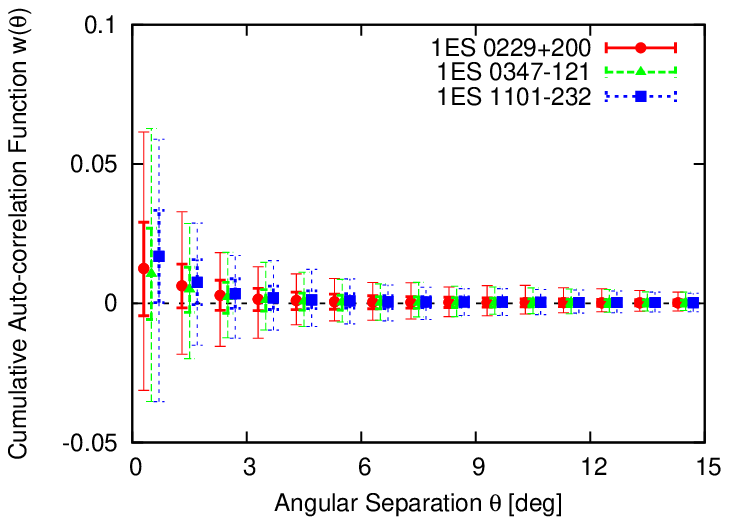}
\includegraphics[clip,width=0.48\linewidth]{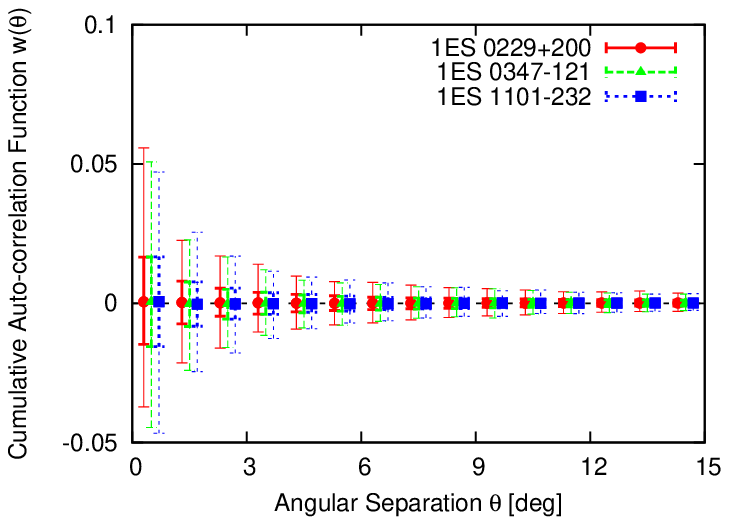}
\caption{Same as figure \ref{fig:ac20}, but with $E_{\rm c} = 10^{19}$ eV.}
\label{fig:ac19}
\end{figure*}

On the other hand, in the case of $E_{\rm c} = 10^{19}$ eV shown in Figure \ref{fig:ac19}, no significant anisotropy appears above $10^{19}$ eV as expected from the CR-luminosity constraints. Note that the Bethe-Heitler process is the main provider of electromagnetic particles to produce the observed VHE gamma rays, so the UHECR-induced cascade model does not necessarily require the acceleration of protons up to $10^{20}$ eV. In the low redshift universe, the energy-loss length of UHECR protons due to Bethe-Heitler pair production is minimized at $\approx 1$ Gpc and $E_p\approx 10^{19}$ eV. However, UHECRs with the energies of $\ll 10^{19}$ eV can be easily deflected by EGMFs surrounding their sources and intervening magnetic fields, and therefore it is unlikely that lower-energy CRs remain beamed for the requirement of the UHECR-induced cascade model.

\begin{figure}
\includegraphics[clip,width=\linewidth]{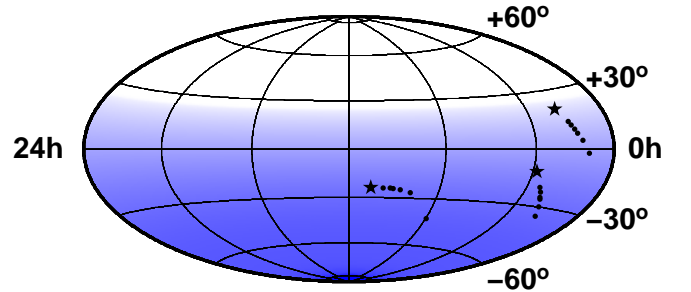}
\caption{Positions of known extreme HBLs, 1ES 0229+200, 1ES 0347-121, and 1ES 1101-232 ({\it stars}), and the arrival directions of UHE protons emitted from these sources with $10^{19.5}$, $10^{19.4}$, $10^{19.3}$, $10^{19.2}$, $10^{19.1}$, $10^{19.0}$ eV ({\it points}; near to far from each source) in equatorial coordinates with the exposure of the PAO ({\it shade}), which is larger at denser points.}
\label{fig:gmf}
\end{figure}

In order to compare the simulation results to the observational data, it is also important to consider the effect of the GMF. The GMF modifies the arrival directions of UHECRs by smearing out or sometimes focusing an event cluster produced by a powerful UHECR source. Figure \ref{fig:gmf} shows the arrival directions of UHECRs from the three EHBLs with $10^{19.5}$, $10^{19.4}$, $\cdots$, $10^{19}$ eV (near to far from the sources indicated by stars) in equatorial coordinates after the modifications by the \citet{AlvarezMuniz:2001vf} GMF model. The relative exposure of the PAO is indicated in color, which is larger at a more southerly direction. It is found that all three EHBLs are in the aperture of the PAO, though the exposure of 1ES 0229+200 is significantly reduced by being near the edge of PAO's field of view. Note that protons above $10^{19.5}$ eV do not have to be considered for these EHBLs because of their energy-loss by photomeson production in a CMB field. The arrival directions of UHECRs deviate significantly from the positions of the sources and move in the directions where the PAO exposure is larger. Hence, the significance of the anisotropic signals caused by the EHBLs is expected to be enhanced. Although the modifications depend on an adopted GMF model, this tendency is unchanged even if recent GMF models \citep[e.g.,][]{Pshirkov2011ApJ738p192,Jansson2012ApJ757p14} are applied. This is because UHECR arrival directions are most modified by the magnetic field in the vicinity of the solar system, which is modeled similarly in all the GMF models.

\begin{figure*}
\includegraphics[clip,width=0.48\linewidth]{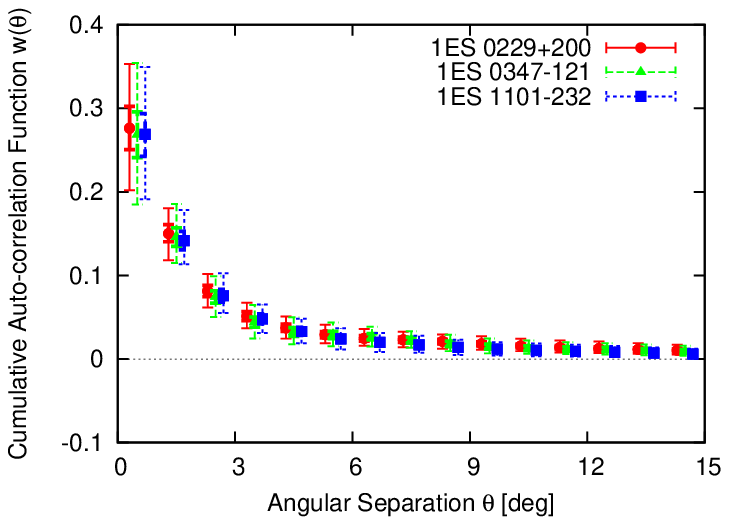}
\includegraphics[clip,width=0.48\linewidth]{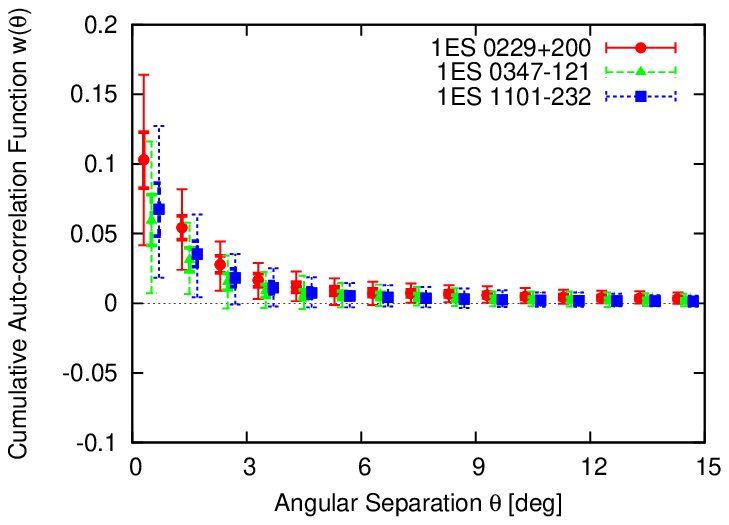} \\
\includegraphics[clip,width=0.48\linewidth]{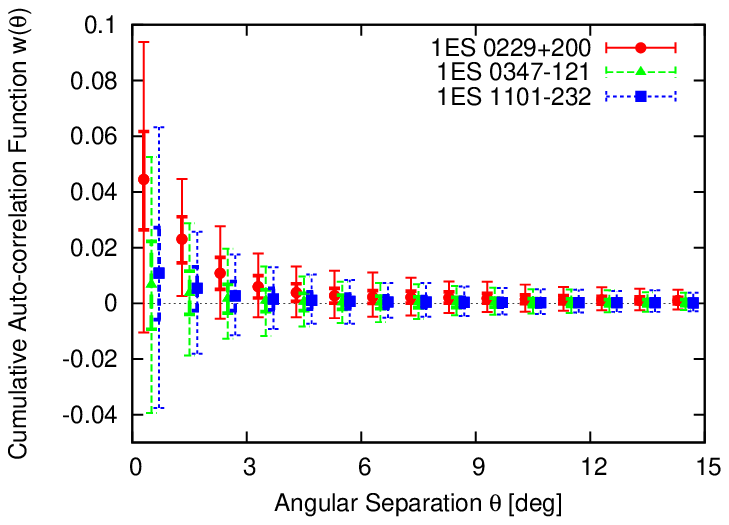}
\includegraphics[clip,width=0.48\linewidth]{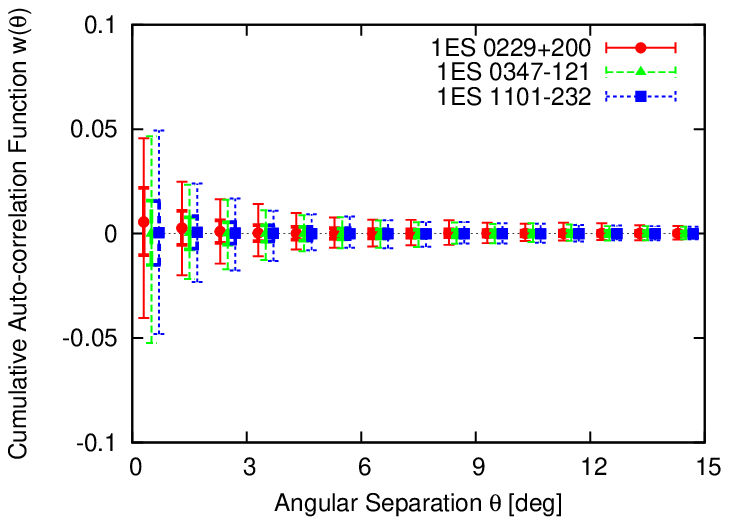}
\caption{Same as Figure 3, except here the aperture geometry of PAO with total exposure of $31645$ km$^2$ yr sr is used to simulate the observational situation of PAO at 2013. The CR spectra are assumed to have $E_{\rm c} = 10^{20}$ eV ({\it upper}) and $E_{\rm c} = 10^{19}$ eV ({\it lower}) with $s = 2.0$ ({\it left}) and $2.6$ ({\it right}). The GMF is taken into account while the local magnetic field is not considered.}
\label{fig:acgmf}
\end{figure*}

Figure \ref{fig:acgmf} shows the cumulative angular auto-correlation functions considered in figures \ref{fig:ac20} and \ref{fig:ac19}, but with the modifications of the arrival directions of UHECRs by the GMF. We can find the two competing effects of the GMF. The modification of the arrival directions of UHECRs to lower declination increases the number of arriving UHECRs, enhancing the signals of anisotropy at small angular scales. This appears mainly for 1ES 0229+200 because the change of the aperture in the directions of UHECRs is large. Remarkably, the significance of the anisotropic signals is more than 3 sigmas even for $E_{\rm c} = 10^{19}$ eV if $s = 2.0$. On the other hand, it also somewhat smears the clustering of events due to the deflections of UHECRs depending on their energies, which is reflected in the absolute values of cumulative auto-correlation functions.

\begin{figure*}
\includegraphics[clip,width=0.48\linewidth]{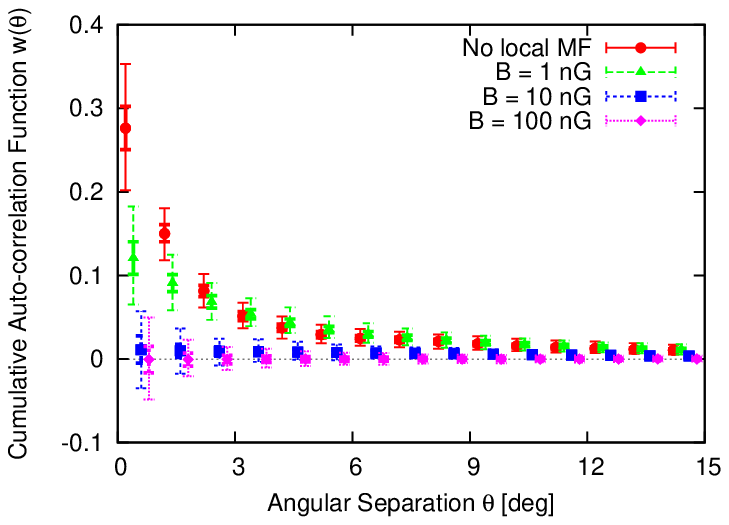}
\includegraphics[clip,width=0.48\linewidth]{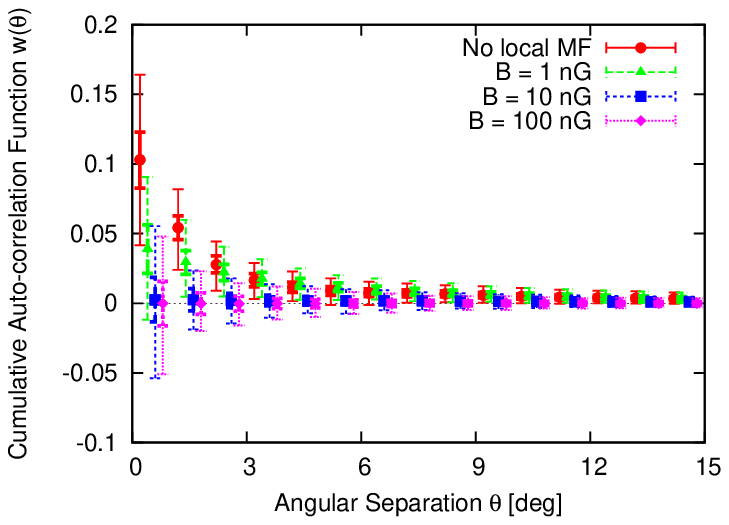}
\caption{Dependence of the cumulative angular auto-correlation function of the arrival direction distribution of UHECRs on the strength of the EGMF surrounding the Milky Way $B_{\rm LEG}$ for the case of 1ES 0229+200. The aperture geometry of PAO with total exposure of $31645$ km$^2$ yr sr is applied. The cutoff energy of the CR spectrum is assumed to be $E_{\rm c} = 10^{20}$ eV. Its spectral index is $s = 2.0$ ({\it left}) and $2.6$ ({\it right}). The GMF is taken into account, with values of the EGMF as noted in the legend.}
\label{fig:1es0229E200}
\end{figure*}

The LEGMF surrounding the Milky Way can reduce the level of the anisotropy without decreasing the $\gamma$-ray flux predicted in the UHECR-induced cascade model because UHECRs providing electromagnetic particles still propagate rectilinearly over almost all the propagation path. Figure \ref{fig:1es0229E200} shows the dependence of the cumulative angular auto-correlation function on the strength of the LEGMF $B_{\rm LEG}$ for the arrival direction distribution of UHECRs consisting of isotropic background and the contribution of 1ES 0229+200. The cutoff energy of the CR spectrum is set to $E_{\rm c} = 10^{20}$ eV and the GMF is taken into account, as noted in the legend. The auto-correlation functions in the no-magnetic field case are the same as those in the upper panels of Figure \ref{fig:acgmf}. If the Milky Way is embedded in a magnetized structure, the anisotropic signal is highly suppressed. Nevertheless, the significance of the anisotropy is still $>3\sigma$ if the strength of the LEGMF is $B_{\rm LEG} \lesssim$ a few nG. The LEGMF with $B_{\rm LEG} = 1$ nG did not cause a large difference from the case with no local magnetic field in the discussion in Section \ref{sec:ul}, but it produces a significant difference here. This is because the deviation of the arrival directions of UHECRs from a source is increased by the effects of the GMF. Since the deflections of protons with $E_p\approx 10^{19}$ eV reach $20^{\circ}$, this enhancement affects the arrival direction distribution of the UHECRs at the earth significantly. If the strength of the LEGMF approaches $10$ nG, UHECRs from the EHBL are so isotropized that the arrival direction distributions of UHECRs are consistent with an isotropic distribution. These conclusions are unchanged for sources with $2\lesssim s \lesssim 2.6$.

\begin{figure*}
\includegraphics[clip,width=0.48\linewidth]{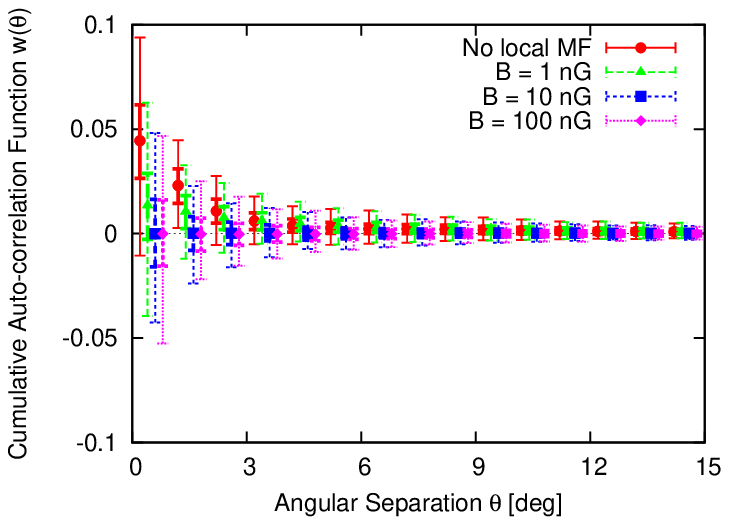}
\includegraphics[clip,width=0.48\linewidth]{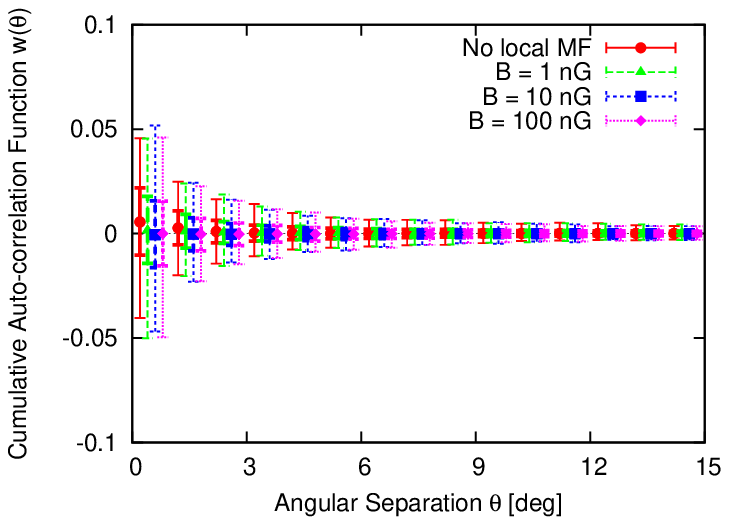}
\caption{Same as figure \ref{fig:1es0229E200}, but $E_{\rm c} = 10^{19}$ eV.} 
\label{fig:1es0229E190}
\end{figure*}

If the cutoff energy of protons is lower, that is, $E_c =10^{19}$ eV, the strength of the LEGMF to achieve the observed isotropy is lower, as shown in Figure \ref{fig:1es0229E190}. Even in the case of 1ES 0229+200, only which has the significance of anisotropy more than 3 sigmas for $B_{\rm LEG} = 0$, $B_{\rm LEG} = 1$ nG is enough to realize arrival distribution consistent with isotropy.

\begin{figure*}
\includegraphics[clip,width=0.48\linewidth]{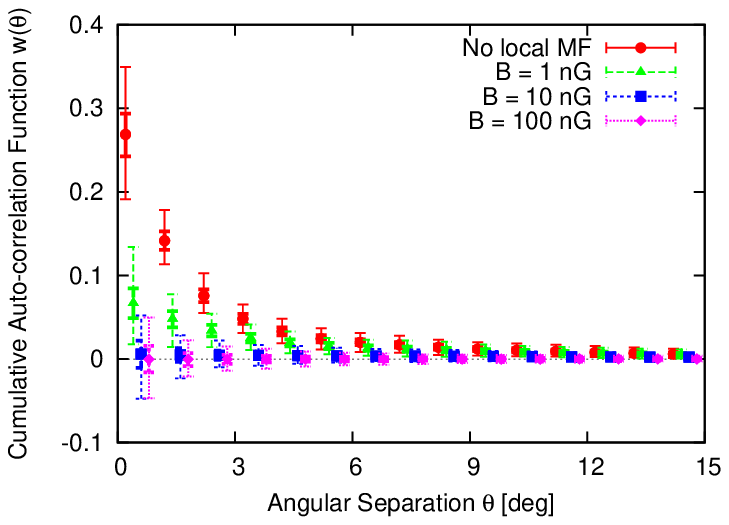}
\includegraphics[clip,width=0.48\linewidth]{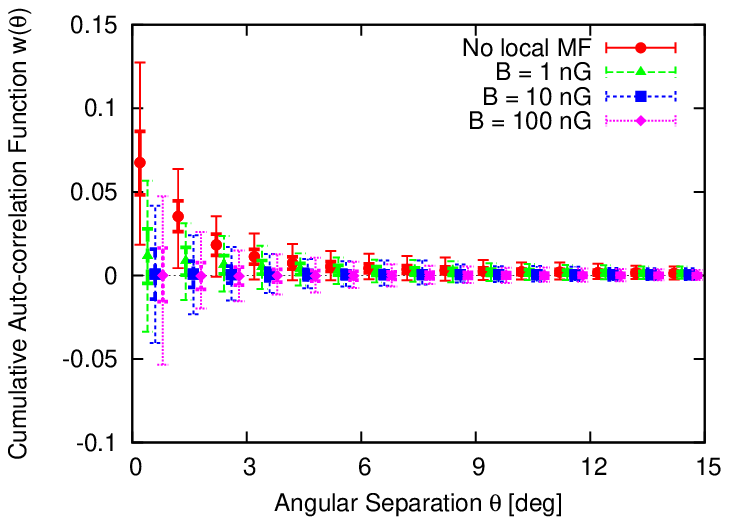}
\caption{Dependence of the cumulative angular auto-correlation function of the arrival direction distribution of UHECRs consisting of isotropic background and the contribution of 1ES 1101-232 on the strength of the local magnetic field $B_{\rm LEG}$. The aperture geometry of PAO with total exposure of $31645$ km$^2$ yr sr is applied to mock the observational situation of PAO in 2013. The cutoff energy is $E_c = 10^{20}$ eV. Two spectral indices, i.e., $s = 2.0$ ({\it left}) and $s = 2.6$ ({\it right}), are treated.The GMF is taken into account, with values of the EGMF as noted in the legend.} 
\label{fig:1es1101}
\end{figure*}

\begin{figure*}
\includegraphics[clip,width=0.48\linewidth]{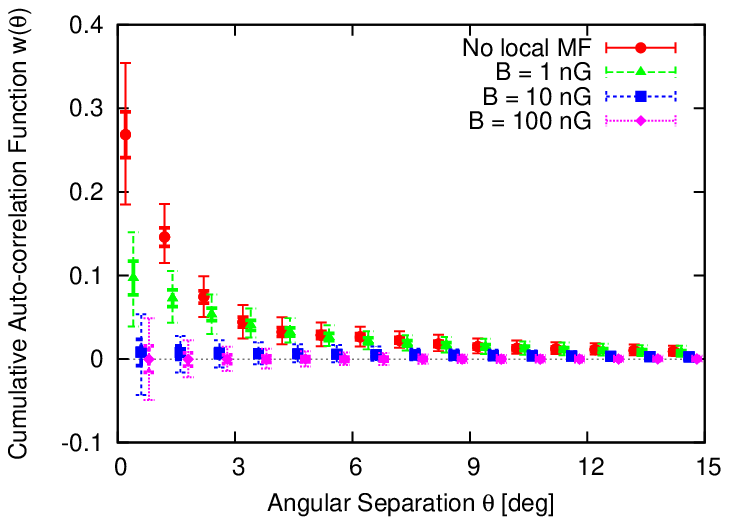}
\includegraphics[clip,width=0.48\linewidth]{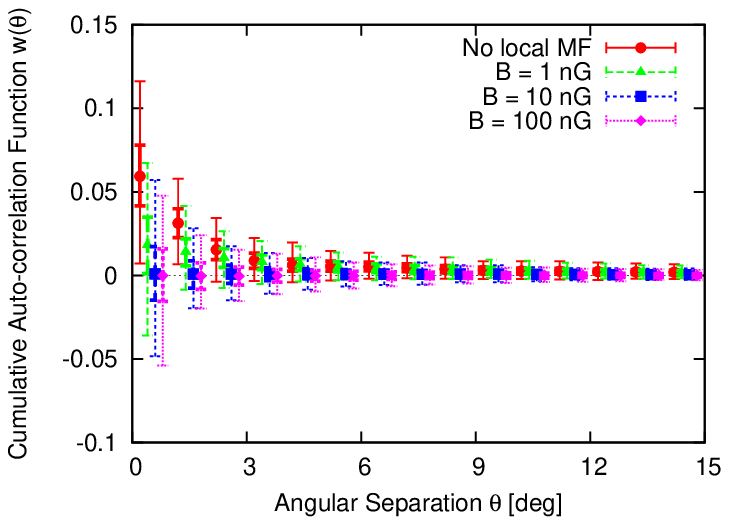}
\caption{Same as figure \ref{fig:1es1101}, but for 1ES 0347-121.}
\label{fig:1es0347}
\end{figure*}

The same discussion can be applied to the other two EHBLs. Figure \ref{fig:1es1101} shows the same as Figure \ref{fig:1es0229E200}, but for 1ES 1101-232. Significant anisotropy appears at $3~\sigma$ level under the LEGMF with $B_{\rm LEG} \lesssim$ a few nG as for 1ES 0229+200 in the hard-spectrum cases. On the other hand, in the steep-spectrum cases, $B_{\rm LEG} = 1$ nG is enough to smear out the anisotropy. These results are quantitatively the same also for 1ES 0347-121, shown in Figure \ref{fig:1es0347}.

At present, no significant small-scale anisotropy has been detected in the arrival direction distributions of UHECRs above $10^{19}$ eV. This isotropy gives interesting constraints on the UHECR-induced cascade model. Based on the discussion in this section, 1ES 0229+200 produces the strongest anisotropy, and therefore provides the strongest constraints among the three. The consistency of the UHECR-induced cascade model holds when 1) the Milky Way is surrounded by a strongly magnetized structure ($B_{\rm LEG} \gtrsim$ a few nG) even if the EHBLs accelerate protons up to the highest energies $E_{\rm c} \sim 10^{20}$ eV, or 2) the strength of the magnetic field around the Milky Way is $B_{\rm LEG} \gtrsim 1$ nG for the hard-spectrum($s = 2.0$) case and lower magnetic field is allowed for the steep-spectrum ($s = 2.6$) cases if the cutoff energy of protons is as low as $E_{\rm c} \sim 10^{19}$ eV, or 3) the maximum acceleration energy of protons is lower than $\sim 10^{19}$ eV, which may indicate that these objects do not accelerate the bulk of the UHECR, depending on composition. Note that $B_{\rm LEG} > 10$ nG is inferred on the assumption that Centaurus A produces UHECRs and explains the local excess in the direction of this object under proton-dominated composition \citep{Yuksel2012ApJ758p16}. 

\section{Discussion} \label{sec:discussion}

We have estimated the general upper limits of CR luminosity by  focusing on the most powerful source in the UHECR sky in Section \ref{sec:ul}. In fact, these are also the upper bounds of CR luminosity even if several powerful sources are taken into account simultaneously. Imagine, for instance, the case that several powerful sources are located at the same redshift. The strongest source contributes to a certain fraction of arriving CRs. The second strongest source contributes also to a certain (but smaller) fraction of arriving CRs. Since the total number of CRs is constant for a fixed exposure, the number of isotropic background events is reduced by the contribution of the second source. The contribution of other focused sources also reduce the number of isotropic background events. As a result, the maximum luminosity to produce significant anisotropy becomes lower, which indicates that the upper limit of the CR luminosity of the strongest source is smaller than that when only a single source is taken into account. Thus, the upper limits estimated in Section \ref{sec:ul} provide robust upper limits of CR luminosity of powerful UHECR emitters at each redshift.

The CR luminosity of individual sources has been constrained by exclusively focusing on gamma-ray components in the framework of the UHECR-induced cascade model. Representative EHBLs, such as EHBLs treated in this paper, require an integrated CR luminosity $L_{\rm CR}^{\rm iso} \sim 10^{45}$ -- $10^{46}$ erg s$^{-1}$ \citep{Essey2010APh33p81,Essey2011ApJ731p51,Murase2012ApJ749p63,Razzaque2012ApJ745p196}, which corresponds to the differential CR luminosity $L_{\rm UHECR,MS}^{\rm iso} \sim 10^{44}$ -- $10^{45}$ erg s$^{-1}$. Also, \citet{Supanitsky2013JCAP12p023} and \citet{Anjos2014JCAP07p049} apply the UHECR-induced cascade model to nearby ($z \gtrsim 0.01$) potential UHECR emitters and provide the bounds of CR luminosities from $\gamma$-ray data. We stress that our anisotropy constraints to the CR luminosity are more stringent as long as $B_{\rm LEG} \lesssim 10$ nG or, in some cases, even if $B_{\rm LEG} \sim 100$ nG. Despite the uncertainty of magnetic fields in extragalactic space, the anisotropy of UHECRs provides strong bounds on the power of UHECR emission.

We have considered only steady sources of UHECRs. However, some UHECR source candidates, such as gamma-ray bursts \citep[GRBs; e.g.,][]{Waxman:1995vg,Vietri1995ApJ453p883} including low-luminosity classes such as trans-relativistic supernovae \citep[e.g.,][]{Murase2006ApJ651L5,Murase2008PRD78p023005}, active galactic nuclei \citep[AGN; e.g.,][]{Biermann1987ApJ322p643,Takahara1990PTP83p1071,Rachen1993AA272p161,Norman1995ApJ454p60,Farrar:2008ex,Dermer2009NJPh11p5016,Pe'er2009PRD80p123018,Takami2011APh34p749,Murase2012ApJ749p63}---especially in flaring phases, and young neutron stars \citep[e.g.,][]{Blasi2000ApJ533L123,Arons2003ApJ589p871,Murase2009PRD79p103001,Kotera2011PhRvD84p023002,Fang2012ApJ750p118} are transient phenomena. Note that in the context of CR astronomy transient phenomena are defined as ones in which the activity lifetime is shorter than the arrival time delay of CRs. The CR flux of a transient source is the CR energy input divided by time dispersion due to cosmic magnetic fields, which is comparable with the time delay of the CRs. Since the time delay is $\propto E_p^{-2}$, the ratio of the source number densities of UHECRs with $10^{19}$ eV and $6 \times 10^{19}$ eV is $\sim 40$ \citep[e.g.,][]{MiraldaEscude1996ApJ462L59,Murase2009ApJ690L14,Takami2009Aph30p306,Takami2012ApJ748p9}. This ratio is consistent with that between $n_s \sim 10^{-4}$~Mpc$^{-3}$ for $E > 6 \times 10^{19}$ eV and $n_s \gtrsim 8 \times 10^{-4}$~Mpc$^{-3}$ for $E > 10^{19}$ eV indicated by the numerical value of equation (\ref{eq:ns19}) (see also \citet{Takami2009Aph30p306} for the earlier estimation of $n_s$ for $E > 10^{19}$ eV). Therefore, if $L_{\rm UHECR,MS}^{\rm iso,ul}(z = 0.01) \lesssim 10^{41}$ erg s$^{-1}$ is established, the difference of the apparent source number densities may indicate the transient generation of UHECRs. This constraint is achievable in the near future because we are likely to live in a filamentary structure \citep{Klypin2003ApJ596p19}, whose magnetic field is indicated to be $\sim 10$ nG by simulations \citep[e.g.,][]{Ryu2008Science320p909}. It was earlier proposed that the difference of the apparent number density of UHECR sources estimated from the anisotropy of UHECRs is a hint of  transient generation of UHECRs \citep{Takami2012ApJ748p9}.

The latter part of this study is motivated by the UHECR-induced cascade model, in which protons must propagate quasi-rectilinearly in voids.  Note that the UHECR-induced cascade model requires $B_{\rm IGV} \lesssim 10^{-14}$ G for electron-positron pairs not to be deflected by void EGMFs, which is consistent with the effective magnetic field $B_{\rm eff} \lesssim 10^{-10}$ G for a coherence length $\lambda_{\rm eff} \approx 0.1$--$1$ Mpc, which is the conservative requirement for anisotropy of UHECRs from blazars to be meaningful (see equation~12). Under these conditions, the effects of the LEGMF surrounding the Milky Way has been examined. The effective magnetic field consists of magnetic fields in filamentary structures and clusters of galaxies. \citet{Takami2012ApJ748p9} estimates the effective magnetic field as $B_{\rm eff} \lambda_{\rm eff}^{1/2} \sim 1$ nG Mpc$^{-1/2}$ following numerical simulations of these structures by \citet{Das2008ApJ682p29}. In this estimation, they adopted the magnetic strength of $\sim 10$ nG and volume-filling factor of the magnetized region of $\sim 0.01$ for filamentary structures, and the magnetic strength of $\sim 0.3 \mu$G and volume-filling factor of $\sim 10^{-4}$ for clusters of galaxies. The coherent length of both magnetic fields is assumed to be 100 kpc. These two structures almost equally contribute to the estimation above. However, in fact, these quantities are still under debate, A smoothed particle hydrodynamical simulation obtains a much smaller volume-filling factor than that of \citet{Das2008ApJ682p29} \citep{Dolag2005JCAP01p009}. In order to realize the requirement of the UHECR-induced cascade model, $B_{\rm eff} \lesssim 10^{-10}$ G for $\lambda \sim 0.1$--1 Mpc, the magnetic strength, coherent length, and volume-filling fraction of these structures are a few orders of magnitude smaller than those assumed in the above estimation.

Alternatively, analytic estimation of the magnetic scattering of UHECRs reveals possible parameter space in which the effective scattering optical depth of UHECRs with $\sim 10^{19}$ eV in extragalactic space is unity or less \citep{Kotera2008PRD77p123003}. UHECRs' trajectories cannot intercept more than a few MSs, implying $\pi R^2 n_{\rm MS} d \lesssim3$, or the number density of MSs $n_{\rm MS} \lesssim 3\times 10^{-4}$ Mpc$^{-3} (R / 2~{\rm ~Mpc})^{-2}$ for the typical distance  $d\lesssim 1$ Gpc of known EHBLs.  This is a factor $\sim 10$ smaller than the density of typical spirals \citep[e.g.,][]{Binney1998GalAstron}, but could be consistent with all blazar sources and all star-forming galaxies like the Milky Way being surrounded by MSs. This is because only a fraction $\sim 10$--20\% of BL Lac objects are weakly variable and may be explained by the UHECR-induced cascade model. Furthermore, there would be a detection preference for distant BL Lac objects whose gamma-rays are formed by UHECR-induced cascades rather than made at the source \citep{Takami2013ApJ771L32}.  Thus, the requirement of the quasi-rectilinear propagation may be compatible with the existence of magnetic fields around the Milky Way and blazar sources. An important study for VHE telescopes, including the Cherenkov Telescope Array, is the variability properties of blazars as a function of redshift. Future observations of these extragalactic magnetic structures, for example, with the Square Kilometer Array, are also essential for testing the UHECR-induced cascade model and the validity of our limits.

\section{Summary} \label{sec:summary}

By means of numerical simulations, we have studied anisotropy in the arrival direction distribution of $> 10^{19}$ eV CRs arriving from steady powerful CR sources. The strong isotropy at $\sim 10^{19}$ eV imposes upper bounds on the CR luminosity of a steady powerful UHECR source, depending on the source redshift and EGMF. We have derived constraints for both current and future observations. This isotropy constraint is stronger than the CR luminosity implied by $\gamma$-ray observations, in the framework of the UHECR-induced cascade model. The isotropic UHECR sky also restricts the existence of typical sources in the local universe and also provides a lower limit on the local source number density of UHECRs with $\sim 10^{19}$ eV. This lower limit on the apparent source number density is much larger than the inferred source number density of UHECRs above $\sim 6 \times 10^{19}$ eV, which may indicate that UHECRs are generated transiently. Note that the isotropy at $\sim 10^{19}$ eV is difficult to reconcile by assuming heavy ions, because composition measurements, depending on hadronic interaction models, imply light composition below $\sim 10^{19}$ eV.

We also examined UHECR anisotropy produced by the EHBLs 1ES 0229+200, 1ES 0347-121 and 1ES 1101-232 in the UHECR-induced cascade model, including effects of the GMF and EGMF. The UHECR-induced model proposed by Essey \& Kusenko (2010) has several attractive features, not least of which is to explain the weakly variable class of blazars, especially the EHBLs considered here, and to account for unusual hard spectral features that appear when deabsorbing high-energy spectra of high-synchrotron-peaked BL Lac objects for a wide range of EBL models. If there were no cosmic magnetic fields, EHBLs would produce strong anisotropy in the UHECR sky at $\sim 10^{19}$ eV, which should be already detected by the PAO unless the CR proton spectra of the EHBLs end well below $10^{19}$ eV. This would rule out the UHECR-induced cascade model under the hypothesis that blazars and their off-axis counterparts are sources of UHECRs. One possible way to maintain this model is if the observed isotropy at $\sim 10^{19}$ eV results from combined effects of a local extragalactic magnetic field (LEGMF) cosmic structure with $B_{\rm LEG}\sim 10$--$100$ nG within which the Milky Way is embedded, in addition to deflections made by the GMF. Measurements of CR anisotropy indicate that the UHECR-induced cascade model works in the cases that 1) the Milky Way is surrounded by a strongly magnetized structure ($B_{\rm LEG} \gtrsim$ a few nG) if EHBLs accelerate protons up to the highest energies, 2) the strength of the LEGMF is $B_{\rm LEG} \gtrsim 1$ nG for the hard-spectrum case and lower magnetic fields are allowed for the steep-spectrum cases, or 3) a lower intrinsic maximum energy of protons, $\lesssim 10^{19}$ eV, is required. The nearly rectilinear propagation of UHECR protons required in the UHECR-induced cascade model is marginally compatible with the assumption that the Milky Way, all star-forming galaxies, and UHECR sources are surrounded by the magnetized structures hypothesized here, depending on the fraction of BL Lac sources that display UHECR components, and observational selection biases favoring detection of blazars with UHECR components (see Section \ref{sec:discussion}). If the UHECR-induced cascade model operates, the required intrinsic CR luminosity has to be 10-100 times larger, depending on the magnetic structure around the sources \citep{Murase2012ApJ749p63}, it could be possible that nearby bright radio galaxies show anisotropy of the arrival directions of $\sim 10^{19}$ eV CRs.  Additional UHECR array exposure will strengthen luminosity limits or detect anisotropy or clustering in UHECR arrival directions and help identify the sources of the UHECRs or, if not, could reveal clues to the existence of an extended magnetic medium in the LEGMF.

\begin{acknowledgements}
We are grateful to Soebur Razzaque, Fumio Takahara, and Eli Waxman for useful discussion.We especially thank John Beacom for stimulating comments on local extragalactic magnetic fields. We also thank an anonymous referee for fruitful comments. This work is supported by Japan Society for the Promotion of Science KAKENHI 24$\cdot$9375 (H. T.), NASA through Hubble Fellowship Grant No. 51310.01 awarded by the STScI, which is operated by the Association of Universities for Research in Astronomy, Inc., for NASA, under Contract No. NAS 5-26555 (K.M.). The work of C.D. is supported by the Chief of Naval Research.
\end{acknowledgements}

\bibliographystyle{apj.bst}

\end{document}